\documentclass[pra,aps,preprint,amsmath,amssymb,showkeys]{revtex4}
\usepackage{graphicx}
\usepackage{epsfig}
\usepackage{epstopdf}
\include{graphics}
\DeclareMathOperator{\sech}{sech}
\DeclareMathOperator{\sgn}{sgn}

\begin{document}

\title{Collision-induced amplitude dynamics of fast 2D solitons in saturable nonlinear media with weak nonlinear loss}

\author{Quan M. Nguyen$^{1}$}
\email[Corresponding author. E-mail address: ]{quannm@hcmiu.edu.vn}

\author{Toan T. Huynh$^{2,3}$}

\affiliation{$^{1}$Department of Mathematics, International University, 
Vietnam National University, Ho Chi Minh City, Vietnam}

\affiliation{$^{2}$Department of Mathematics, University of Science, 
Vietnam National University, Ho Chi Minh City, Vietnam}

\affiliation{$^{3}$Department of Mathematics, University of Medicine and Pharmacy at Ho Chi Minh City, Ho Chi Minh City, Vietnam}

\date{\today}

\begin{abstract}
We study the amplitude dynamics of two-dimensional (2D) solitons in a fast collision described by the coupled nonlinear Schr\"odinger equations with a saturable nonlinearity and weak nonlinear loss. We extend the perturbative technique for calculating the collision-induced dynamics of two one-dimensional (1D) solitons to derive the theoretical expression for the collision-induced amplitude dynamics in a fast collision of two 2D solitons. Our  perturbative approach is based on two major steps. The first step is the standard adiabatic perturbation for the calculations on the energy balance of perturbed solitons and the second step, which is the crucial one, is for the analysis of the collision-induced change in the envelope of the perturbed 2D soliton. Furthermore, we also present the dependence of the collision-induced amplitude shift on the angle of the two 2D colliding-solitons. 
In addition, we show that the current perturbative technique
can be simply applied to study the collision-induced amplitude shift in a fast collision of
two perturbed 1D solitons.
Our analytic calculations are confirmed by numerical simulations with the corresponding coupled nonlinear Schr\"odinger equations in the presence of the cubic loss and in the presence of the quintic loss.
\end{abstract}


\maketitle


\section{Introduction}
\label{Introduction}

Solitons are stable shape preserving solitary waves propagating in
nonlinear dispersive media. Solitons have attracted
considerable attentions in recent years due to the broadened applications of solitons in modern science
 \cite{Ablowitz_2011,Tao_2006,Agrawal_2003,Agrawal_2013,Malomed_2011,Malomed_2019}.
In fact, solitons appear in a variety of fields, including optics, nanophotonics, condensed matter physics \cite{Ablowitz_2011, Agrawal_2013},
and plasma physics \cite{Horton_1996}. In optics, the one-dimensional soliton propagation is stable and can be described by the nonlinear Schr\"odinger (NLS) model \cite{Agrawal_2013}. 
Due to the stability of the temporal NLS solitons, they can be used as bits of information in the optical fiber transmission technology \cite{Agrawal_2013}.
However, 2D solitons are generally unstable in nonlinear optical media \cite{Malomed_2011}.
In particular, 2D optical solitons do not propagate in uniform cubic (Kerr) nonlinear media because of the catastrophic beam collapse at high powers \cite{Berge,Eisenberg}.
There have been several investigations to achieve the stabilization of 2D solitons \cite{Malomed_2002,Yang_2010,Ablowitz_2012,Malomed_2004a}.
It was shown that 2D solitons can be stabilized in a layered structure with sign-alternating Kerr nonlinearity
\cite{Malomed_2002}.
They also can exist and be stabilized in Kerr nonlinear optical media with an external potential
\cite{Yang_2010,Ablowitz_2012,Malomed_2004a}.
Recently, the existence and stability of 2D optical solitons in saturable nonlinear media are subjects of continuously renewed interest to achieve the stable transmission of light beams at high velocity. The theoretical analyses of the condition for the existence of 2D and 3D solitons in saturable media were developed in Ref. \cite{Malomed_2004}. 
Saturable nonlinearities have been observed in many nonlinear materials including photorefractive materials such as LiNbO$_3$ \cite{Agrawal_2003,Weilnau}.
The 2D solitons can exist in photorefractive crystals due to the relatively slow nonlinear response of these materials. When the light goes through these media, the refractive index changes and the material might force the light to remain confined in its self-generated waveguide.
As a result, the light can propagate without changing the shape. Additionally, it was shown that 2D solitons can be stabilized in the nonlinear media where the cubic domains are embedded into materials with saturable nonlinearities \cite{Torner_2010}. 
In such optical media, the soliton propagation can also be described by (2+1)-dimensional ((2+1)D) NLS equation with a saturable nonlinearity \cite{Agrawal_2003,Weilnau,CalVar2013}.

One of the most fundamental properties of ideal solitons is their shape-preserving property in a soliton collision, that is, a soliton collision is elastic \cite{ZK1965}. In optics, the collisions of sequences of solitons are very frequent \cite{Ablowitz_2011, Agrawal_2013}.
 Therefore, the collisions of two and many 1D solitons have been intensively investigated in several studies, for example, see Refs. \cite{Com_Math_Phys_2009,Soneson_2004,CP2005,Perelman11, PNC2010,PNG2014,Peleg2019} and references therein. More specifically, in Refs. \cite{PNC2010, PNG2014}, the authors studied the 1D soliton collision-induced amplitude dynamics in the presence of the cubic loss and the generic nonlinear loss. In optics, the nonlinear loss arises due to multiphoton absorption (MPA) or gain/loss saturation in silicon media \cite{Boyd-2008,PNG2014}. MPA has been received considerable attention in recent years due to the importance of
MPA in silicon nanowaveguides, which are expected to play
a crucial role in optical processing applications in optoelectronic devices, including pulse switching and compression,
wavelength conversion, regeneration, etc. \cite{PNC2010, PNG2014, Peleg2019, Boyd-2008, Husko2009, Husko2013, Loon2018}. It has been uncovered that the presence of weak nonlinear loss leads to an additional downshift of the soliton amplitude in a fast collision of two 1D solitons \cite{PNC2010, PNG2014}.
The analytic expressions for the amplitude shift in two-soliton collisions, which is described by the (1+1)D NLS model, in the presence of weak cubic loss, which can be a result of TPA or gain and loss saturation, were already found in Refs. \cite{PNC2010,NH2019} and in the presence of the weak $(2m+1)-$order loss, for any $m \geq 1$, were found in Ref. \cite{PNG2014}. In the previous studies for 1D soliton collision-induced change in the four parameters of solitons \cite{Soneson_2004,CP2005,PNC2010,PNG2014}, the perturbative techniques were based on the projections of the total collision-induced change in the soliton envelope on the four localized eigenmodes of the linear operator $\hat{L}$ describing small perturbations about the
fundamental NLS soliton, which was derived by Kaup in the 1990s \cite{Kaup1990,Kaup1991}. However, in this original perturbation theory, the soliton solution of the unperturbed model, which is the $\sech$-form soliton, was used for the calculations on the dynamics of perturbed 1D solitons.
Consequently, it is very hard to apply a similar technique for studying the effects of small perturbations on the interactions of solitons in higher dimensions, in which the unperturbed equations are nonintegrable. One needs to develop a new approach for studying the soliton collision-induced dynamics in the presence of nonlinear dissipation in higher dimensions instead of using the Kaup's perturbation theory. It is worthy to note that the collision-induced corrections to solitons amplitudes were investigated in Ref. \cite{Dyachenko_1989} in the framework of {\it unperturbed} nonintegrable wave models. However, the study for the soliton amplitude dynamics in the nonintegrable wave models with {\it nonlinear dissipation} has not been explored. So far, to the best of our knowledge, the study for the collision-induced amplitude dynamics of 2D solitons in saturable nonlinear media in the presence of nonlinear dissipation is a long standing open problem.

In this work, this important and challenging problem will be addressed. We study fast collisions between two 2D solitons in weakly perturbed nonlinear optical media.
The dynamics of the collision is described by the systems of coupled (2+1)D NLS equations
with the saturable nonlinearity, which are nonintegrable models, in the presence of the generic weak $(2m+1)$-order of the nonlinear loss, for any $m \geq 1$. We derive the analytic expression for the amplitude dynamics of a 2D single-soliton and, particularly, for the collision-induced amplitude dynamics in a collision of two fast 2D solitons in the presence of weak nonlinear loss.
For the aforementioned purposes, we develop a perturbative method for perturbed 2D solitons. Our perturbative method significantly extends the perturbative technique in Refs. \cite{Soneson_2004,CP2005,PNC2010,PNG2014} for calculating the effects of weak perturbations on fast collisions between two 1D solitons of the NLS equation and the recent perturbative method presented in Ref. \cite{PNH2017b} for calculating the collision-induced amplitude dynamics of two 1D pulses in perturbed linear waveguides. The crucial points in the current perturbative approach are the uses of the solution of the {\it perturbed} NLS model instead of the unperturbed NLS model and the single soliton dynamics in calculating the total collision-induced change in the soliton envelope. These are the key improvements compared to the perturbation techniques for studying the perturbed 1D solitons presented in Refs. \cite{Soneson_2004,CP2005,PNC2010,PNG2014}.
More specifically, our perturbative approach is based on a procedure of two steps. The first step is for calculations on the energy balance of perturbed solitons based on the perturbed solution and a standard adiabatic perturbation
theory for solitons. The second step, which plays a crucial role for our perturbative approach, is for calculations on the collision-induced change in the soliton envelope and a technical approximation of integrals based on the assumption of a fast and complete collision. We verify the analytic expressions by the numerical simulations with the corresponding (2+1)D NLS models with the cubic loss $(m=1)$ and with the quintic loss $(m=2)$.
Additionally, we also demonstrate that the current perturbative approach 
can be simply applied to calculate the collision-induced amplitude
shift in a fast collision of two 1D solitons for a large class of perturbed (1+1)D NLS equations in a straightforward manner. As a concrete example, 
we use the current perturbative technique to
derive the expression for the collision-induced amplitude shift in a fast collision of
two 1D solitons of (1+1)D cubic NLS model in the presence of the delayed Raman response.

The rest of the paper is organized as follows.
In sections \ref{analytic_1a}, \ref{analytic_1b}, and \ref{analytic_1c}, we first study the dynamics of a single-soliton propagation in saturable nonlinear optical media in the presence of the generic weak nonlinear loss.
Then, we use the perturbative technique to calculate the collision-induced amplitude dynamics in a fast collision of two 2D solitons. 
The analytic predictions will be validated by simulations in section \ref{simulation}.
Section \ref{conclusion} is reserved for conclusions.
In Appendix \ref{Appen_NLS2}, we demonstrate the robustness and the simplicity of the current perturbative method for other perturbed soliton equations.

\section{Collision-induced amplitude dynamics of two 2D solitons}
\label{analytic}
\subsection{The perturbed coupled (2+1)D NLS equations and the ideal 2D solitons}
\label{analytic_1a}

We consider fast collisions between two 2D solitons propagating in saturable nonlinear optical media in the presence of the weak $(2m+1)$-order of the nonlinear loss, for any $m \ge 1$. The dynamics of the collision is described by the system of coupled NLS equations as follows \cite{Agrawal_2003,Weilnau, CalVar2013,PNG2014}:
\begin{eqnarray}&&
\!\!\!\!\!\!\!\!\!\!\!\!\!\!
i\partial_{z}\psi_{j} + \Delta_{\bot}\psi_{j} + \frac{\alpha(|\psi_{j}|^{2} + 
|\psi_{l}|^{2})}{1+(|\psi_{j}|^{2} + |\psi_{l}|^{2})/I_{0}}\psi_{j}
\!\!\!\!\!\!\!\!\!\!\!\!\!\! 
\nonumber \\&&
\!\!\!\!\!\!\!\!\!\!\!\!\!\!
 =  -i\epsilon_{2m+1}|\psi_{j}|^{2m}\psi_{j}
 -i\epsilon_{2m+1}\sum\limits_{k = 1}^{m} b_{k,m}|\psi_{l}|^{2k}|\psi_{j}|^{2(m-k)}\psi_{j},
 \!\!\!\!\!\!\!\!\!\!\!\!\!\!
\label{NLS_coll1}
\end{eqnarray}
where $b_{k,m} = \frac{m!(m+1)!}{(k!)^{2}(m+1-k)!(m-k)!}$ \cite{PNG2014}, 
$1\le j,l \le 2$ and $j\ne l$,
$\Delta_{\bot} = \partial^{2}_{x} + \partial^{2}_{y}$ is the transverse Laplace operator,
$\psi_{j}$ is the envelope of soliton $j$,
$x$ and $y$ are the spatial coordinates, $z$ is the propagation distance, $\alpha$ is the strength of the nonlinearity, $I_{0}$ is the saturation parameter,
and $\epsilon_{2m+1}$, which satisfies $0<\epsilon_{2m+1} \ll 1$, is the $(2m+1)$-order of the nonlinear loss coefficient \cite{PNG2014,dimensionless}.
On the left-hand side (LHS) of equation (\ref{NLS_coll1}), the second term corresponds to the second-order dispersion and the third term represents the effects of the saturable nonlinearity. On the right-hand side (RHS) of equation (\ref{NLS_coll1}), the first and second terms describe the effects of intra-beam and inter-beam interaction due to the $(2m+1)$-order of the nonlinear loss, respectively. Note that the effect of the generic nonlinear loss on interaction of 1D cubic NLS solitons was uncovered in Ref. \cite{PNG2014}. In addition, its specific cases, the weak cubic loss $(m=1)$ and the weak quintic loss $(m=2)$, were also investigated in Refs. \cite{PNC2010,Peleg2019} and \cite{Husko2009, Husko2013}, respectively.

We first discuss the form of the single ideal 2D soliton $j$ which is the fundamental solution of the following unperturbed model
\cite{Agrawal_2003,Torner_2010}:
\begin{eqnarray}&&
\!\!\!\!\!\!\!
i\partial_{z}\psi_{j} + \Delta_{\bot}\psi_{j}
 + \frac{\alpha |\psi_{j}|^{2}}{1+ |\psi_{j}|^{2}/I_{0}}\psi_{j}
 = 0.
\label{NLS_single1a}
\end{eqnarray}
The soliton solution of equation (\ref{NLS_single1a}) with the velocity vector $\textbf{d}_{j} = (d_{j1},d_{j2})$ can be found in the form: 
\begin{eqnarray}&&
\!\!\!\!\!\!\!
\tilde\psi_{j0}(x,y,z) = U_{j}(X_{j},Y_{j})\exp(i\mu_{j} z)
\exp\left[i\alpha_{j} + i\chi_{j}(\tilde X_{j},\tilde Y_{j})\right],
\label{NLS_sol}
\end{eqnarray} 
where 
$X_{j} = x - x_{j0} - d_{j1}z$, $Y_{j} = y - y_{j0} - d_{j2}z$,
$\tilde X_{j} = x - x_{j0} - \tilde d_{j1}z$, $\tilde Y_{j} = y - y_{j0} - \tilde d_{j2}z$,
$\chi_{j}= \tilde d_{j1}\tilde X_{j} + \tilde d_{j2} \tilde Y_{j}$, $\tilde d_{j1} = d_{j1}/2$, $\tilde d_{j2} = d_{j2}/2$, $(x_{j0},y_{j0})$ is the initial position of soliton $j$, $\alpha_{j}$ is related to the phase,
$d_{j1}$ and $d_{j2}$ correspond to the velocity components in the $x$ and $y$ directions, respectively,
$\mu_{j}$ is the propagation constant, and $U_{j}$ is the localized real-valued amplitude function. 
From equations \eqref{NLS_single1a} and \eqref{NLS_sol}, it can be shown that the function $U_{j}$ satisfies the following elliptic equation \cite{Agrawal_2003,Torner_2003}:
\begin{eqnarray}&&
\!\!\!\!\!\!\!
\Delta_{\bot}U_{j} + \frac{\alpha U_{j}^{3}}{1+ U_{j}^{2}/I_{0}} = \mu_{j}U_{j}.
\label{NLS_sol2}
\end{eqnarray}

\subsection{The 2D soliton dynamics of the single-soliton propagation}
\label{analytic_1b}

Next, we investigate the effects of the $(2m+1)$-order of the nonlinear loss on the single-soliton propagation described by the following perturbed equation:
\begin{eqnarray}&&
\!\!\!\!\!\!\!
i\partial_{z}\psi_{j} + \Delta_{\bot}\psi_{j}
 + \frac{\alpha |\psi_{j}|^{2}}{1+ |\psi_{j}|^{2}/I_{0}}\psi_{j}
 =   -i\epsilon_{2m+1}|\psi_{j}|^{2m}\psi_{j}.
\label{NLS_single1}
\end{eqnarray}
By using an energy balance calculation for equation (\ref{NLS_single1}), it implies:
\begin{eqnarray}&&
\!\!\!\!\!\!\!\!\!\!\!\!
\partial_{z}\int_{-\infty}^{\infty}\int_{-\infty}^{\infty} \!\!\!\!\!\!|\psi_{j}|^{2}dxdy\!=
-2\epsilon_{2m+1}\int_{-\infty}^{\infty}\int_{-\infty}^{\infty} \!\!\!\!\!\!|\psi_{j}|^{2m+2}dxdy.
\label{NLS_single2}
\end{eqnarray}       
We assume that the initial envelopes of the 2D solitons can be expressed in the 
general form  
\begin{eqnarray} &&
\psi_{j0}(x,y,0)= A_{j}(0)\tilde\psi_{j0}(x,y,0),
\label{IC_single1}
\end{eqnarray} 
where $A_{j}(0)$ is the initial amplitude parameter, $\tilde\psi_{j0}(x,y,0)$ is the fundamental soliton solution of equation (\ref{NLS_single1a}), that is, $\tilde\psi_{j0}(x,y,0)$ is given by equation (\ref{NLS_sol}), and $j=1,2$. We note that for an initial envelope of the unperturbed soliton solution, one can define $A_{j}(0)=1$, that is $\psi_{j0}(x,y,0)= \tilde\psi_{j0}(x,y,0)$. In the presence of the nonlinear loss, we look for the solution of equation (\ref{NLS_single1}) in the form of
\begin{eqnarray} &&
\psi_{j0}(x,y,z)= A_{j}(z)\tilde\psi_{j0}(x,y,z),
\label{sol_single1}
\end{eqnarray} 
where $A_{j}(z)$, $0 < A_{j}(z) < A_j(0)$, is the amplitude parameter taking into account of the effects of nonlinear loss for $z>0$, and $\tilde\psi_{j0}(x,y,z)$ is given by equation (\ref{NLS_sol}).
We substitute the relation for $\psi_{j0}(x,y,z)$ into the equation (\ref{NLS_single2}) and apply the standard adiabatic perturbation
theory for the NLS soliton \cite{Hasegawa95}. It then yields:
\begin{eqnarray} &&
\!\!\!\!\!\!\!
\frac{d}{dz}\left[ I_{2,j}(z)A_{j}^{2}(z) \right] = - 2\epsilon_{2m+1}I_{2m+2,j}(z)A_{j}^{2m+2}(z),
\!\!\!\!\!\!\!
\label{NLS_ODE1}
\end{eqnarray}
where
 $$I_{2,j}(z)=\int_{-\infty}^{\infty} \int_{-\infty}^{\infty} |\tilde\psi_{j0}(x,y,z)|^{2}dxdy = \int_{-\infty}^{\infty} \int_{-\infty}^{\infty} U_{j}^{2}dxdy,$$
and 
$$I_{2m+2,j}(z)=\int_{-\infty}^{\infty}\int_{-\infty}^{\infty} |\tilde\psi_{j0}(x,y,z)|^{2m+2}dxdy = \int_{-\infty}^{\infty} \int_{-\infty}^{\infty} U_{j}^{2m+2}dxdy.$$
By the definition of $U_{j}$, one can obtain that $I_{2,j}(z)$ and $I_{2m+2,j}(z)$ are constants. Solving equation (\ref{NLS_ODE1}) on the interval $[0,z]$, it implies the equation for the amplitude dynamics of a single soliton as follows: 
\begin{eqnarray} &&
\!\!\!\!\!\!\!
A_{j}(z)= \frac{A_{j}(0)}{ \left[1 + 2m\epsilon_{2m+1}I_{2m+2,j0}/I_{2,j0} A_{j}^{2m}(0)z\right]^{1/(2m)} },
\!\!\!\!\!\!\!
\label{NLS_Ampli1}
\end{eqnarray}
where $I_{2,j0}=I_{2,j}(0)$ and $I_{2m+2,j0}=I_{2m+2,j}(0)$.

Equation (\ref{NLS_Ampli1}) describes the effects of the nonlinear loss on the amplitude parameter of a single 2D soliton. It also shows that in the leading order of perturbation effects, the amplitude $A_1(z)$ decays at order proportional to $\mathcal{O}(z^{-\frac{1}{2m}})$. 

\subsection{The collision-induced amplitude dynamics of two 2D solitons}
\label{analytic_1c}

We now study the collision-induced amplitude dynamics in a fast two-soliton collision described by equation (\ref{NLS_coll1}). 
For this purpose, we assume two solitons are well-separated at the initial propagation distance $z=0$ and at the final propagation distance $z=z_f$ for a complete collision.
By deriving the energy balance of equation (\ref{NLS_coll1}), one then obtains:
\begin{eqnarray}&&
\!\!\!\!\!\!\!
\partial_{z}\int_{-\infty}^{\infty}\int_{-\infty}^{\infty} |\psi_{j}|^{2} dxdy=
-2\epsilon_{2m+1}\int_{-\infty}^{\infty}\int_{-\infty}^{\infty} |\psi_{j}|^{2m+2} dxdy
-2\epsilon_{2m+1}\sum\limits_{k = 1}^{m} b_{k,m}J^{(j,l)}_{k,m},
\!\!\!\!\!\!\!\!
\label{NLS_coll2}
\end{eqnarray}
where $J^{(j,l)}_{k,m} = \int_{-\infty}^{\infty}\int_{-\infty}^{\infty}
 |\psi_{l}|^{2k}|\psi_{j}|^{2(m-k) + 2} dxdy$.
Based on the perturbative calculation approach in Refs. \cite{PNC2010,NH2019,CP2005}, it is useful to look for 
the solution of equation (\ref{NLS_coll1}) in the form:
\begin{eqnarray}&&
\!\!\!\!\!\!\!
\psi_{j}(x,y,z)=\psi_{j0}(x,y,z)+\phi_{j}(x,y,z), 
\label{NLS_coll3}
\end{eqnarray}       
where $\psi_{j0}$ is the single-soliton propagation solution of equation (\ref{NLS_single1}) and $\phi_{j}$ describes a small correction to $\psi_{j0}$, i.e., the correction is solely due to collision effects.
We substitute the relation (\ref{NLS_coll3}) into equation (\ref{NLS_coll2}) and take into account only leading-order effects, that is, the effects of order of $\epsilon_{2m+1}$. 
Therefore, based on the standard adiabatic perturbation
theory for the NLS soliton \cite{Hasegawa95}, the terms containing $\phi_{j}$ on the RHS of the resulting equation can be neglected. It then leads to the following differential equation for soliton 1:
\begin{eqnarray}&&
\!\!\!\!\!\!\!
\partial_{z}\int_{-\infty}^{\infty}\int_{-\infty}^{\infty} |\psi_{1}|^{2} dxdy=
-2\epsilon_{2m+1}\int_{-\infty}^{\infty}\int_{-\infty}^{\infty} |\psi_{10}|^{2m+2} dxdy
-2\epsilon_{2m+1}
\sum\limits_{k = 1}^{m} b_{k,m}K_{k,m},
\!\!\!\!\!\!\!\!
\label{NLS_coll4}
\end{eqnarray}
where $K_{k,m}=\int_{-\infty}^{\infty}\int_{-\infty}^{\infty}
|\psi_{20}|^{2k}|\psi_{10}|^{2(m-k) + 2}dxdy$.
Equation \eqref{NLS_coll4} represents the energy balance for soliton 1. The last term on the RHS of equation \eqref{NLS_coll4} is responsible for the contribution of the interaction term during the collision. We note that when $\epsilon_{2m+1}=0$ then equation \eqref{NLS_coll4} becomes a conservation law for the energy and the calculations to obtain the equation for soliton 2 are the same.
From equations (\ref{NLS_single2}) and (\ref{NLS_coll4}) and noting that $\psi_{j0}$ satisfies equation (\ref{NLS_single2}), one then obtains the energy balance equation for soliton 1 via the use of the {\it perturbed} single-soliton propagation solution as follows:
\begin{eqnarray}&&
\!\!\!\!\!\!\!\!
\partial_{z}\int_{-\infty}^{\infty}\int_{-\infty}^{\infty} |\psi_{1}|^{2} dxdy=
\partial_{z}\int_{-\infty}^{\infty}\int_{-\infty}^{\infty} |\psi_{10}|^{2} dxdy
-2\epsilon_{2m+1}
\sum\limits_{k = 1}^{m} b_{k,m}K_{k,m}.
\!\!\!\!\!\!\!\!
\label{NLS_coll5}
\end{eqnarray}
In a {\it fast} collision, the collision takes place in a small interval $[z_{c}-\Delta z_{c}, z_{c}+\Delta z_{c}]$ around $z_{c}$, where $z_{c}$ is the collision distance, which is the distance at which the maxima of $|\psi_{j}(x,y,z)|$ coincide at the same point $(x_0,y_0)$, and $\Delta z_{c}$ is the distance along which the envelopes of the colliding solitons overlap ($\Delta z_{c} \ll 1$). Integrating over $z$ of equation (\ref{NLS_coll5}), it implies:
\begin{eqnarray}&&
\!\!\!\!\!\!\!\!\!\!\!\!\!\!
\int_{z_{c}-\Delta z_{c}}^{z_{c}+\Delta z_{c}}
\partial_{z}\int_{-\infty}^{\infty}\int_{-\infty}^{\infty} |\psi_{1}|^{2} dxdydz
= 
 \int_{z_{c}-\Delta z_{c}}^{z_{c}+\Delta z_{c}}
\partial_{z}\int_{-\infty}^{\infty}\int_{-\infty}^{\infty} |\psi_{10}|^{2} dxdydz
-2\epsilon_{2m+1} 
\sum\limits_{k = 1}^{m} b_{k,m}L_{k,m},
\!\!\!\!\!\!\!\!\!\!\!\!\!\!
\nonumber \\
\label{NLS_coll6}
\end{eqnarray}
where $L_{k,m}=\int_{z_{c}-\Delta z_{c}}^{z_{c}+\Delta z_{c}}
\int_{-\infty}^{\infty}\int_{-\infty}^{\infty}
|\psi_{20}|^{2k}|\psi_{10}|^{2(m-k) + 2}
dxdydz$.

Let us derive the expression for the collision-induced amplitude shift in a fast two-soliton collision from equation (\ref{NLS_coll6}).
These calculations are based on the approximations on the total collision-induced change in the soliton envelope via the use of the {\it perturbed} single-soliton solution and the conserved quantity of the unperturbed propagation equation.
Let $\Delta_{1}$ and $\Delta_{10}$ be the integral on the LHS and the first integral on the right-hand side of equation (\ref{NLS_coll6}), respectively. That is, 
\[\Delta_{1} = \int_{z_{c}-\Delta z_{c}}^{z_{c}+\Delta z_{c}}\partial_{z}\int_{-\infty}^{\infty}\int_{-\infty}^{\infty}|\psi_{1}|^{2}dxdydz,\]
and 
\[\Delta_{10} = \int_{z_{c}-\Delta z_{c}}^{z_{c}+\Delta z_{c}}\partial_{z}\int_{-\infty}^{\infty}\int_{-\infty}^{\infty}|\psi_{10}|^{2}dxdydz.\]
The expression for $\Delta_{1}$ can be expressed in the term of the change in the soliton envelope:
\begin{eqnarray}&&
\Delta_{1}=\int_{-\infty}^{\infty}\int_{-\infty}^{\infty}|\psi_{1}(x,y,z_{c}^{+})|^{2}dxdy
- \int_{-\infty}^{\infty}\int_{-\infty}^{\infty}|\psi_{1}(x,y,z_{c}^{-})|^{2}dxdy.
\label{NLS_coll6_A1}
\end{eqnarray}
where $z_{c}^{-}=z_{c}-\Delta z_{c}$ and $z_{c}^{+}=z_{c}+\Delta z_{c}$.
Let $\Psi_{j0}(x,y,z)=U_{j}(x-x_{j0}-d_{j1}z,y-y_{j0}-d_{j2}z)$.
We introduce the following approximation:
\begin{eqnarray}&&
|\psi_{1}(x,y,z_{c}^{+})| = \left(A_{1}(z_{c}^{-}) + \Delta A_{1}^{(s)}(z_{c}) + \Delta A_{1}^{(c)}\right)\Psi_{10}(x,y,z_{c}^{+}),
\label{NLS_coll6_A1_add1}
\end{eqnarray}
where $\Delta A_{1}^{(c)}$ is the total collision-induced amplitude shift of soliton 1, $A_{j}(z_{c}^{-})$ is the limit from the left of $A_{j}(z)$ at $z_{c}$,
and $\Delta A_{1}^{(s)}(z_{c})$ is the amplitude shift of soliton 1 which is due to the single-soliton propagation from $z^{-}_{c} $ to $z^{+}_{c}$.
By the definition of $\psi_{1}$, $\psi_{10}$, and $U_{j}$:
\begin{eqnarray}&&
|\psi_{1}(x,y,z_{c}^{-})| = A_{1}(z_{c}^{-})\Psi_{10}(x,y,z_{c}^{-}).
\label{NLS_coll6_A1_add2}
\end{eqnarray}
Substituting the relations (\ref{NLS_coll6_A1_add1}) and (\ref{NLS_coll6_A1_add2}) into equation (\ref{NLS_coll6_A1}), it yields the following key approximation for the total collision-induced change in the soliton envelope:
\begin{eqnarray}&&
\!\!\!\!\!\!\!
\Delta_{1}=
\left(A_{1}(z_{c}^{-}) + \Delta A_{1}^{(s)}(z_{c}) + \Delta A_{1}^{(c)}\right)^{2}
\int_{-\infty}^{\infty}\int_{-\infty}^{\infty} \Psi_{10}^{2}(x,y,z_{c}^{+})dxdy
\!\!\!\!\!\!\!
\nonumber\\&&
- A_{1}^{2}(z_{c}^{-})\int_{-\infty}^{\infty}\int_{-\infty}^{\infty}\Psi_{10}^{2}(x,y,z_{c}^{-})dxdy.
\!\!\!\!\!\!\!
\label{NLS_coll6_A2}
\end{eqnarray}
We note that $\int_{-\infty}^{\infty}\int_{-\infty}^{\infty} \Psi_{10}^{2}(x,y,z)dxdy$ is
a conserved quantity of the propagation equation (\ref{NLS_single1}) when $\epsilon_{2m+1}=0$.
Therefore, the following relation holds
\begin{eqnarray}&&
\int_{-\infty}^{\infty}\int_{-\infty}^{\infty} \Psi_{10}^{2}(x,y,z)dxdy = I_{2,10}
\label{NLS_coll6_A2_add1}
\end{eqnarray}
for all $z$. Substituting the relation (\ref{NLS_coll6_A2_add1}) into equation (\ref{NLS_coll6_A2}) and taking into account only leading order terms, it implies
\begin{eqnarray}&&
\Delta_{1}= 2A_{1}(z_{c}^{-})\left(\Delta A_{1}^{(c)} + \Delta A_{1}^{(s)}(z_{c}) \right) I_{2,10}.
\label{NLS_coll6_A3}
\end{eqnarray}
On the other hand, $\Delta_{10}$ can be expressed as
\begin{eqnarray}&&
\Delta_{10}=\int_{-\infty}^{\infty}\int_{-\infty}^{\infty}|\psi_{10}(x,y,z_{c}^{+})|^{2}dxdy
- \int_{-\infty}^{\infty}\int_{-\infty}^{\infty}|\psi_{10}(x,y,z_{c}^{-})|^{2}dxdy.
\label{NLS_coll6_A4}
\end{eqnarray}
By the definition of $\psi_{10}$, one can use the approximations $|\psi_{10}(x,y,z_{c}^{-})| = A_{1}(z_{c}^{-})\Psi_{10}(x,y,z_{c}^{-})$
and
$|\psi_{10}(x,y,z_{c}^{+})| = \left(A_{1}(z_{c}^{-}) + \Delta A_{1}^{(s)}(z_{c})\right)\Psi_{10}(x,y,z_{c}^{+}).$
Substituting these relations into equation (\ref{NLS_coll6_A4})
and then expanding the first integrand on the right-hand side while keeping only leading terms, it implies 
\begin{eqnarray}&&
\Delta_{10}= 2A_{1}(z_{c}^{-})\Delta A_{1}^{(s)}(z_{c})I_{2,10}.
\label{NLS_coll6_A5}
\end{eqnarray} 
We substitute equations (\ref{NLS_coll6_A3}) and  (\ref{NLS_coll6_A5}) into equation (\ref{NLS_coll6}).
It arrives at the equation for the collision-induced amplitude dynamics of soliton 1:
\begin{eqnarray}&&
\!\!\!\!\!\!\!
A_{1}(z_{c}^{-})\Delta A_{1}^{(c)} I_{2,10} = -\epsilon_{2m+1}
\sum\limits_{k = 1}^{m} b_{k,m} M_{k,m},
\!\!\!\!\!\!\!\!
\label{NLS_coll7}
\end{eqnarray}
where $M_{k,m} = \int_{z_{c}-\Delta z_{c}}^{z_{c}+\Delta z_{c}}
\int_{-\infty}^{\infty}\int_{-\infty}^{\infty}
A_{2}^{2k}(z)A_{1}^{2(m-k)+2}(z)
U_{2}^{2k}U_{1}^{2(m-k) + 2} dxdydz$,

Finally, we simplify equation (\ref{NLS_coll7}) by integrating $M_{k,m}$ using the {\it decompose} approximation of the integrand based on the assumption of a {\it fast} soliton collision. 
We note that only functions on the right-hand side of equation (\ref{NLS_coll7}) that contain {\it fast} variations in $z$, which are the factors $X_{j}$ and $Y_{j}$, are $U_{1}$ and $U_{2}$. The slow varying amplitudes $A_{1}(z)$ and $A_{2}(z)$ can be approximated by $A_{1}(z_{c}^{-})$ and $A_{2}(z_{c}^{-})$, respectively. Therefore, equation (\ref{NLS_coll7}) can be re-written:
\begin{eqnarray}&&
\!\!\!\!\!\!\!
\Delta A_{1}^{(c)}=
-\epsilon_{2m+1}/ I_{2,10}
\sum\limits_{k = 1}^{m} b_{k,m}A_{2}^{2k}(z_{c}^{-})A_{1}^{2(m-k)+1}(z_{c}^{-})N_{k,m},
\!\!\!\!\!\!\!\!
\label{NLS_coll7b}
\end{eqnarray}
where $N_{k,m} = \int_{z_{c}-\Delta z_{c}}^{z_{c}+\Delta z_{c}}\int_{-\infty}^{\infty}\int_{-\infty}^{\infty}
U_{2}^{2k}U_{1}^{2(m-k) + 2}
dxdydz$.
Since the integrand on the right-hand side of equation (\ref{NLS_coll7b}) is sharply peaked at a small interval about $z_{c}$, we can extend the limits of this integral to $0$ and $z_{f}$. Therefore, it yields 
\begin{eqnarray}&&
\!\!\!\!\!\!\!
\Delta A_{1}^{(c)}=
-\epsilon_{2m+1}/I_{2,10}
\sum\limits_{k = 1}^{m} b_{k,m}A_{2}^{2k}(z_{c}^{-})A_{1}^{2(m-k)+1}(z_{c}^{-})P_{k,m},
\!\!\!\!\!\!\!\!
\label{NLS_coll8}
\end{eqnarray}
where $P_{k,m} = 
\int_{0}^{z_{f}}\int_{-\infty}^{\infty}\int_{-\infty}^{\infty}
U_{2}^{2k}U_{1}^{2(m-k) + 2}dxdydz$. Equation \eqref{NLS_coll8} represents the collision-induced amplitude shift of two fast 2D solitons propagating in optical media with a saturable nonlinearity and the generic weak nonlinear loss. It shows that the collision-induced amplitude dynamics of two fast 2D solitons with nonlinear loss is independent of the phase of the initial solitons.
This behavior is the same as one for fast 1D soliton collisions with cubic loss studied in Ref. \cite{PNC2010}. These observations on the phase-independence in equation (\ref{NLS_coll8}) reveal the interesting difference between the effects of {\it nonlinear dissipation} on the {\it fast} collision-induced amplitude dropdown presented in the current work and of the {\it Raman-induced energy exchange} of 1D soliton studied in Ref. \cite{Antikainen_2012}, where the phase difference between the colliding solitons strongly affects the amplitude of colliding solitons in a {\it slow} collision.

We note that equation (\ref{NLS_coll6_A2}) plays an important role in our analysis. In previous studies for the collision-induced dynamics of two 1D solitons \cite{Soneson_2004,CP2005, PNC2010,PNG2014}, the perturbative method derived by Kaup was mainly based on integrating the correction term $\phi_j$ and the projection of the total collision-induced change in the soliton envelope on the four localized eigenmodes of the linear operation $\hat{L}$ describing small perturbations about the fundamental NLS soliton.
While our approach uses the single perturbed soliton solution of the perturbed NLS equation (\ref{NLS_single1}) and the conserved quantity of the unperturbed propagation equation in calculating the collision-induced change in the soliton envelope. 
Consequently, the current perturbative method can be used to study the effects of weak nonlinear loss on the collision-induced amplitude dynamics of fast 2D solitons of the (2+1)D NLS equations and of other nonintegrable models in 2D and higher spatial dimensions. Moreover, it is worthy to emphasize that the derivation of the LHS of equation \eqref{NLS_coll7} is independent of the type of dissipative perturbations. Therefore, the current perturbative approach can be applied for other types of weak dissipative perturbations, which contribute on the RHS of equation \eqref{NLS_coll7}, in a similar manner.

\section{Numerical simulations}
\label{simulation}
\subsection{Set up the measurements}
\label{simu_mea}

First, let us describe a collision between two solitons as follows.
For simplicity and without loss of generality, we assume the collision occurs at the origin $O(0,0)$ in the $xy$-plane.
Solitons 1 and 2, which are located at $M_{1}(x_{10},y_{10})$ and $M_{2}(x_{20},y_{20})$, respectively, are well-separated at $z=0$.
These two solitons propagate toward $O(0,0)$ with the velocity vectors $\textbf{d}_{1}$ and $\textbf{d}_{2}$.
As a result, the group velocity difference of the colliding-solitons is $\textbf{d} = \textbf{d}_{1} - \textbf{d}_{2}$.
In simulations, $\textbf{d}_{1}$ and $\textbf{d}_{2}$ are chosen as: 
\begin{eqnarray}&&
\!\!\!\!\!\!\!
x_{10}/d_{11} = y_{10}/d_{12}= x_{20}/d_{21}= y_{20}/d_{22}.
\!\!\!\!\!\!\!\!
\label{NLS_num1}
\end{eqnarray}
Therefore, there will be a collision at the origin at the propagation distance $z_{c}=-x_{10}/d_{11}$. 
After the full collision, two solitons continue propagating away from $(0,0)$ and they are thus well-separated at the final propagation distance $z=z_{f}$. Denoting by $\theta$ the collision angle between two colliding-solitons, one can 
determine 
$\cos\theta=\textbf{u}_{1}\cdot \textbf{u}_{2}/(| \textbf{u}_{1} | |\textbf{u}_{2}| ),$
where $\textbf{u}_{1}=\overrightarrow{M_{1}O}$ and $\textbf{u}_{2}=\overrightarrow{M_{2}O}$. 
It then yields
$\cos\theta
=\textbf{d}_{1}\cdot \textbf{d}_{2}/(| \textbf{d}_{1}| | \textbf{d}_{2} |).
$

Second, we define the relative error in the approximation of $\Delta A_{1}^{(c)}$ by
$|\Delta A_{1}^{(c)(num)} - \Delta A_{1}^{(c)(th)}|/|\Delta A_{1}^{(c)(th)}|$, where $\Delta A_{1}^{(c)(th)}$ is calculated from the theoretical prediction of equation (\ref{NLS_coll8}) and  $\Delta A_{1}^{(c)(num)}$ is defined by:
\begin{eqnarray}&&
\!\!\!\!\!\!\!
\Delta A_{1}^{(c)(num)} = A_{1}(z_{c}^{+}) - A_{1}(z_{c}^{-}).
\!\!\!\!\!\!\!\!
\label{NLS_num4}
\end{eqnarray}
In equation (\ref{NLS_num4}), $A_{1}(z_{c}^{-})$ is measured from equation (\ref{NLS_Ampli1})
and $A_{1}(z_{c}^{+})$ is calculated by solving equation (\ref{NLS_ODE1}) with $j=1$ on the interval $[z_{c},z_{f}]$:
\begin{eqnarray} &&
\!\!\!\!\!\!\!
A_{1}(z^{+}_{c})= \frac{A_{1}(z_{f})}{ \left[1 - 2m\epsilon_{2m+1}I_{2m+2,10}/I_{2,10}A_{1}^{2m}(z_{f})(z_{f} - z_{c})\right]^{1/(2m)} },
\!\!\!\!\!\!\!
\label{NLS_Ampli2}
\end{eqnarray}
where $A_{1}(z_{f})$ is measured by simulations of equation (\ref{NLS_coll1}). 
From equation (\ref{NLS_coll8}) the expression for $\Delta A_{1}^{(c)(th)}$ can be written as
\begin{eqnarray}&&
\!\!\!\!\!\!\!
\Delta A_{1}^{(c)(th)}=
-2\epsilon_{3}/ I_{2,10}A_{1}(z_{c}^{-})A_{2}^{2}(z_{c}^{-})P_{1,1}
\!\!\!\!\!\!\!\!
\label{NLS_coll9}
\end{eqnarray}
with cubic loss ($m=1$) and it is 
\begin{eqnarray}&&
\!\!\!\!\!\!\!
\Delta A_{1}^{(c)(th)}=
-\epsilon_{5}/I_{2,10}
\left[ 
6A_{2}^{2}(z_{c}^{-})A_{1}^{3}(z_{c}^{-})P_{1,2} 
+ 3A_{2}^{4}(z_{c}^{-})A_{1}(z_{c}^{-})P_{2,2} 
\right]
\!\!\!\!\!\!\!\!
\label{NLS_coll10}
\end{eqnarray}
with quintic loss ($m=2$).
Additionally, we define the relative error in measuring the soliton patterns at the propagation distance $z$ by $\Big\lVert |\psi^{(th)}| - |\psi^{(num)}|\Big\rVert / \Big\lVert \psi^{(th)}\Big\rVert$,
where $\Big\lVert \psi \Big\rVert = \left[\int_{x_{\min}}^{x_{\max}}\int_{y_{\min}}^{y_{\max}}
|\psi|^{2}dxdy \right]^{1/2}$,
$\psi^{(num)}(x,y,z)$ is measured by simulations, $\psi^{(th)}(x,y,z)$ is the theoretical prediction of the soliton pattern at the propagation distance $z$,
and $[x_{\min}, x_{\max}]\times [y_{\min}, y_{\max}]$ is the computational spatial domain. 
We then define the threshold levels $E_{a}$ and $E_{s}$.
The {\it small error} domain in measuring the analytic prediction of equation (\ref{NLS_coll8}) is determined when the relative errors in measuring $\Delta A_{1}^{(c)}$ and in measuring the soliton patterns at $z=z_{f}$ are less than or equal to $E_{a}$ and $E_{s}$, respectively.
In contrast, the {\it large error} domain in measuring the analytic prediction of equation (\ref{NLS_coll8}) is determined when the relative error in measuring $\Delta A_{1}^{(c)}$ 
is greater than $E_{a}$ or the relative error in measuring the soliton patterns at $z=z_{f}$ is greater than $E_{s}$.
In simulations, $E_{a}=0.1$ and $E_{s} = 0.04$ are used.

Next, we study the dependence of $\Delta A_{1}^{(c)}$ on $\theta$ when the values of $\theta$ are changed over $[0,\pi]$ while the magnitudes of $\textbf{d}_{1}$ and $\textbf{d}_{2}$ are constant. One can define the relative change of $\Delta A_{1}^{(c)}$ due to $\theta$, for $0 \leq \theta \leq \pi$, with respect to $\Delta A_{1,\pi}^{(c)}$ as
\begin{eqnarray} &&
\!\!\!\!\!\!\!
p=[\Delta A_{1}^{(c)} - \Delta A_{1,\pi}^{(c)}]/\Delta A_{1,\pi}^{(c)},
\label{NLS_relp}
\end{eqnarray}
where $\Delta A_{1,\pi}^{(c)}$ is the value of $\Delta A_{1}^{(c)}$ at $\theta=\pi$.
We note that $\Delta A_{1,\pi}^{(c)}$ is the smallest value of $\Delta A_{1}^{(c)}$ over $[0,\pi]$.

To validate equation (\ref{NLS_coll8}), we carry out the simulations with equations (\ref{NLS_coll1}) and (\ref{NLS_single1}) using the split-step Fourier method with the second-order accuracy \cite{Yang_2010}. That is, the errors for numerically solving equations (\ref{NLS_coll1}) and (\ref{NLS_single1}) are of order $\mathcal{O}(h^3)$, where $h=\Delta z$ is the propagation step-size \cite{Yang_2010}. As an example, we present the simulation results of equation (\ref{NLS_coll1}) for $\alpha=1$ and $I_{0} =1$ with $m=1$ and $m=2$. The initial conditions of equation (\ref{NLS_coll1}) are defined from equation (\ref{NLS_sol}) at $z=0$. The ground state $U_{j}$ in equation (\ref{NLS_sol}) at any fixed initial distance $z_{0}\ge 0$ is measured by simulations of equation (\ref{NLS_sol2}) using the accelerated imaginary-time evolution method \cite{Yang_2010,Yang_2008}. To implement simulations with equation (\ref{NLS_sol2}), we use the input function $\tilde U_{j}=\sech\left[(x-x_{j0}-d_{j1}z_{0})^{2} + (y-y_{j0}-d_{j2}z_{0})^{2}\right]$.
Also we use the input power value of the beam $j$ as $P_{j0}=22.5$.
By the simulations of equation (\ref{NLS_sol2}), one can obtain the power value of the soliton solution $j$ as $P_{j}=22.5$ at $\mu_{j}=0.1629$, where 
$P_{j}(\mu_{j})=\int_{x_{\min}}^{x_{\max}}\int_{y_{\min}}^{y_{\max}} U_{j}^{2}(X_{j},Y_{j};\mu_{j})dxdy$.
Additionally, the length of the computational spatial domain is $Lx=Ly=30\pi$ and the number of grid points in $x$-domain and in $y$-domain is $Nx=Ny=2048$. As a result, the spacing of the grid points is $\Delta x=\Delta y = Lx/Nx=0.046$ and $[y_{\min}, y_{\max}]=[x_{\min}, x_{\max}] = [-Lx/2, Lx/2-\Delta x]$. The propagation step-size is $\Delta z=5 \times 10^{-4}$. With this choice of $\Delta z$, the errors for solving equations (\ref{NLS_coll1}) and (\ref{NLS_single1}) are thus in the order of $10^{-10}$. Also, we use the initial phases $\alpha_{j}=0$ and emphasize that the simulation results are independent of the choices of the initial phases $\alpha_{j}$.

\subsection{Simulation results}

\begin{figure}[ptb]
\begin{tabular}{cc}
\epsfxsize=12cm  \epsffile{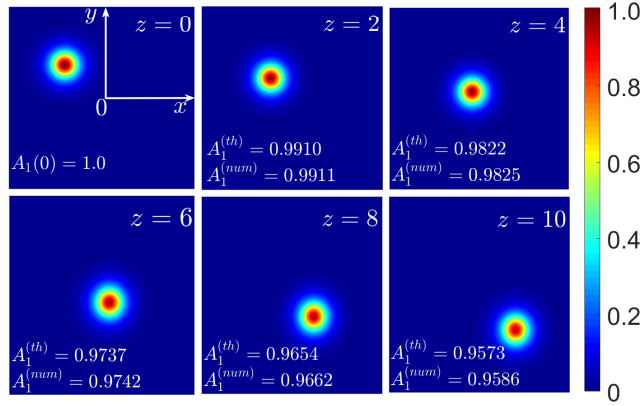} 
\end{tabular}
\caption{(Color online) The initial soliton profile and the evolution of its profiles $|\psi_{1}(x,y,z)|$ obtained by the simulation of the single soliton propagation of equation (\ref{NLS_single1}) with $\epsilon_{3}=0.01$.
}
\label{fig2_add}
\end{figure}

Before validating equation (\ref{NLS_coll8}), we first verify equation (\ref{NLS_Ampli1}) by carrying out simulations for the single-soliton propagation described by equation (\ref{NLS_single1}) for $j=1$ with $m=1$ and $m=2$. The parameters are: $\epsilon_{2m+1}=0.01$, $(x_{10},y_{10})=(-10, 9)$, $\textbf{d}_{1}=(d_{11}, d_{12})=(2, -1.8)$. The final propagation distance is $z_{f}=10$.
Figure \ref{fig2_add} represents the initial soliton profile and the evolution of its profiles $|\psi_{1}(x,y,z)|$ obtained by the simulation of equation (\ref{NLS_single1}) with $\epsilon_{3}=0.01$ at the propagation distances of $z=2, 4, 6, 8, 10$. The soliton profiles are presented using the level colormap. In addition, the amplitude parameters $A_{1}^{(num)}(z)$ and $A_{1}^{(th)}(z)$ are calculated, where $A_{1}^{(num)}(z)$ is measured by the simulation of equation (\ref{NLS_single1}) and $A_{1}^{(th)}(z)$ is calculated from the theoretical prediction with equation (\ref{NLS_Ampli1}).
The agreement between the analytic calculations and the simulation results for $m=1$ is very good. In fact, the relative error in measuring $A_{1}(z)$ for $z \in [0,z_{f}]$, which is defined by $|A_{1}^{(num)}(z)-A_{1}^{(th)}(z)|/A_{1}^{(th)}(z)$, is less than $1.4 \times 10^{-3}$.
The relative error in measuring the soliton patterns over $[0,z_{f}]$ is less than $0.031$. In addition, by implementing the simulation of equation (\ref{NLS_single1}) with $\epsilon_{5}=0.01$ ($m=2$), we obtain the excellent agreement between the simulation results and the analytic calculations of the amplitudes and of the soliton patterns. The relative errors in measuring $A_{1}(z)$ and in measuring the soliton patterns over $[0,z_{f}]$ are less than $1.2 \times 10^{-3}$ and $0.026$, respectively. Moreover, these numerical results also indicate that in average, the amplitude $A_1(z)$ decays at order proportional to $\mathcal{O}(z^{-0.492})$ with $m=1$ and proportional to $\mathcal{O}(z^{-0.246})$  with $m=2$, for $z \in [0,z_{f}]$. These extensive numerical simulations above firmly validate equation (\ref{NLS_Ampli1}) and the amplitude decay of order $\mathcal{O}(z^{-\frac{1}{2m}})$.

\begin{figure}[ptb]
\begin{tabular}{cc}
\epsfxsize=7cm  \epsffile{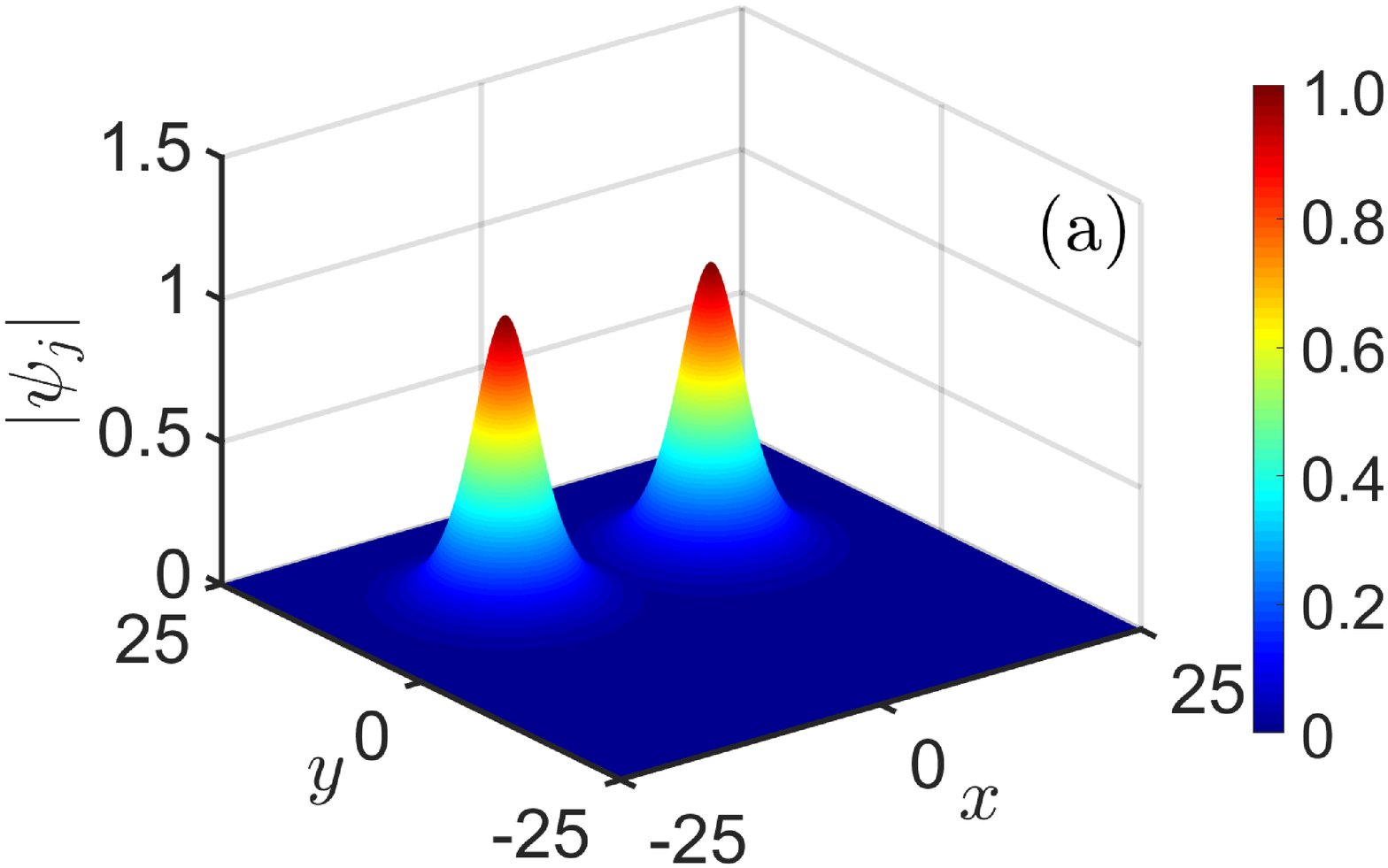} &
\epsfxsize=7cm  \epsffile{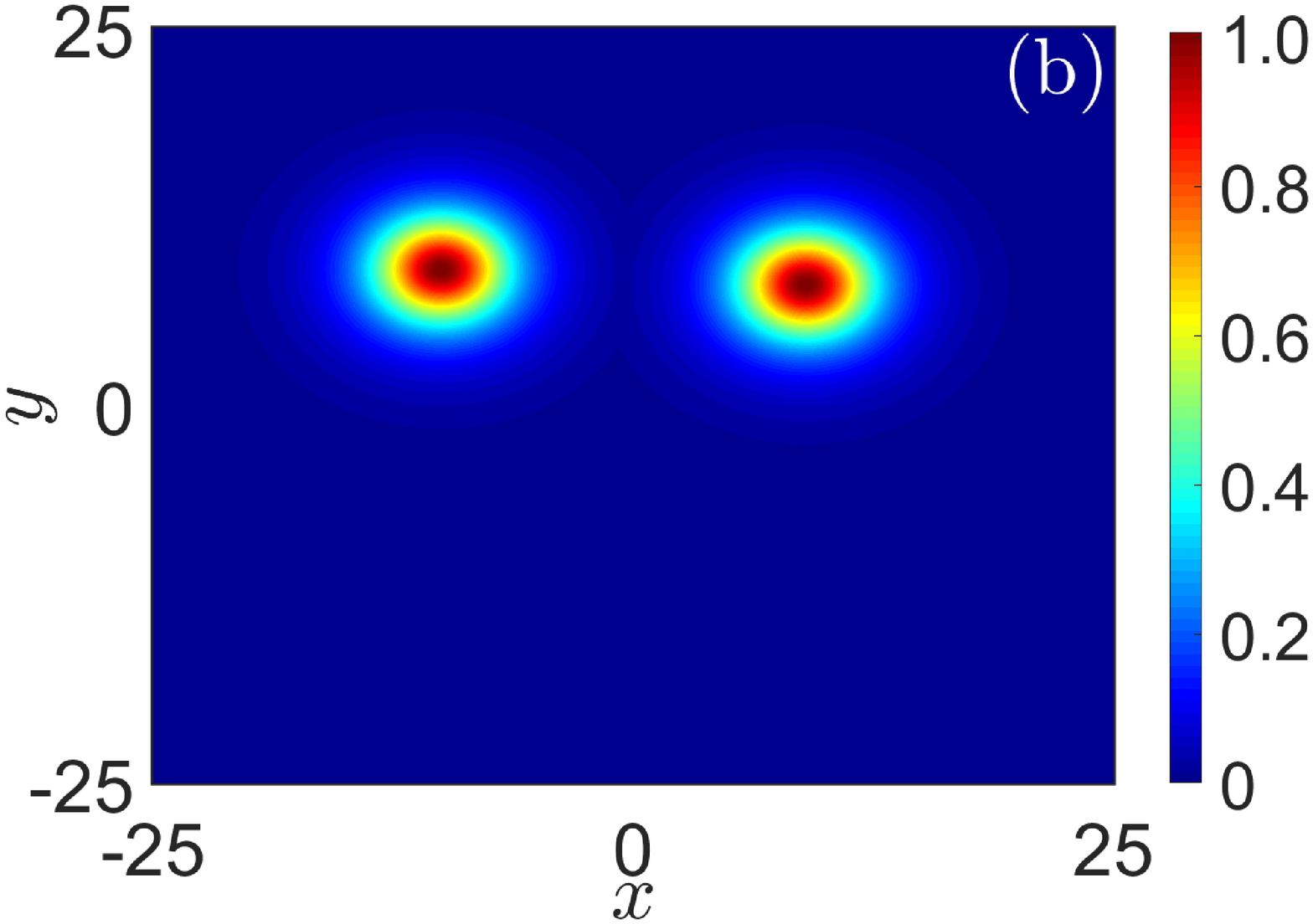} \\
\epsfxsize=7cm  \epsffile{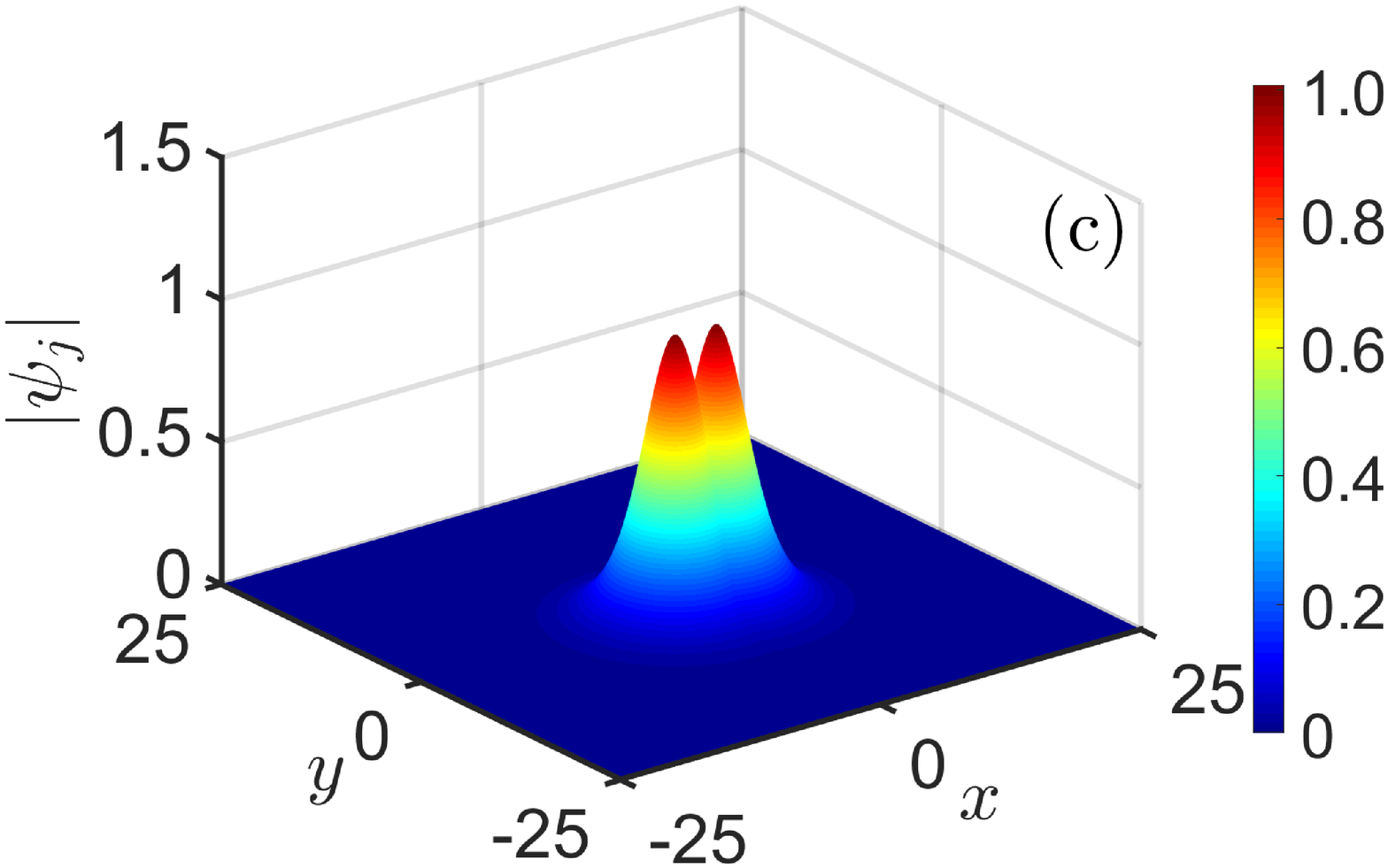} &
\epsfxsize=7cm  \epsffile{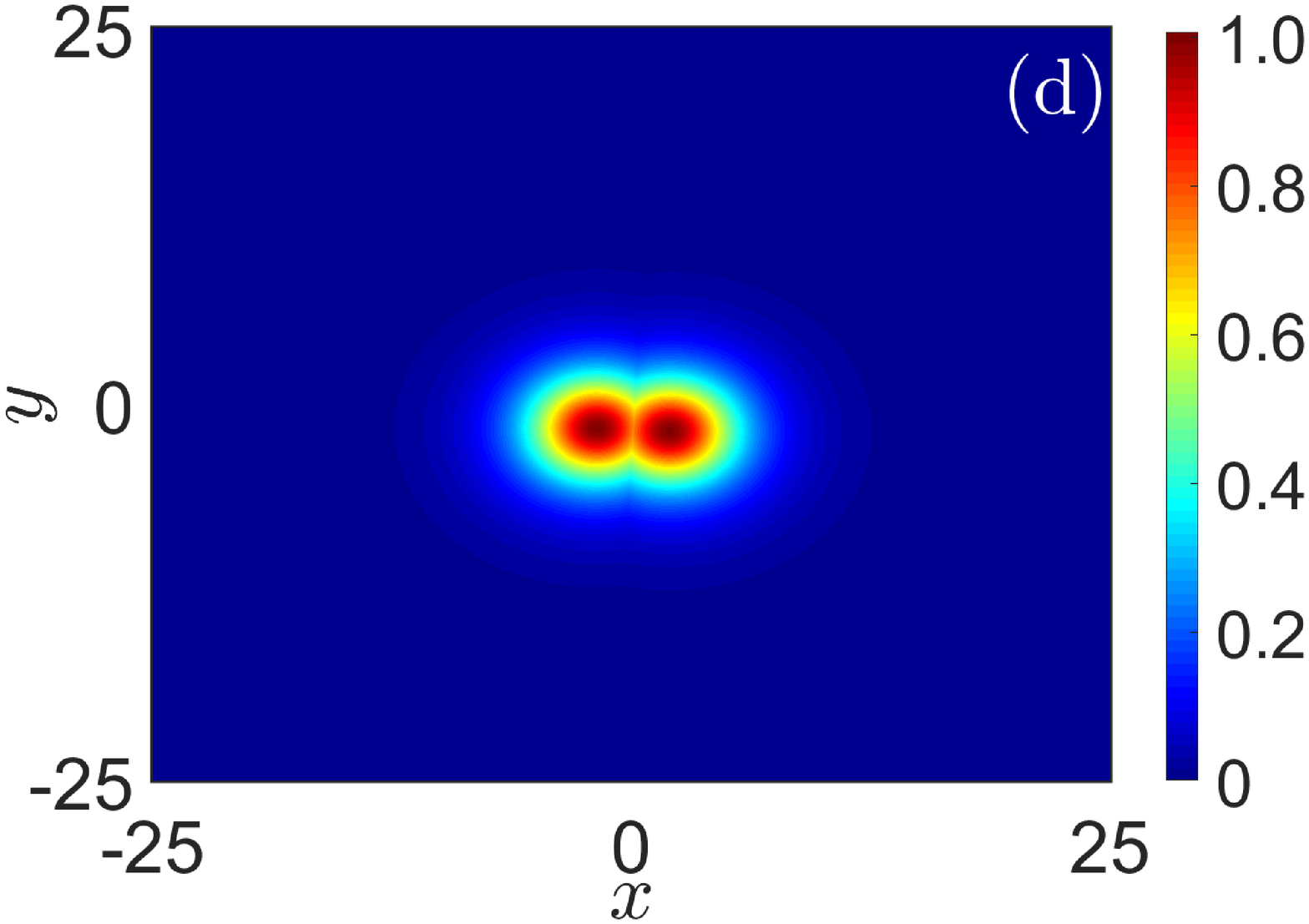} \\
\epsfxsize=7cm  \epsffile{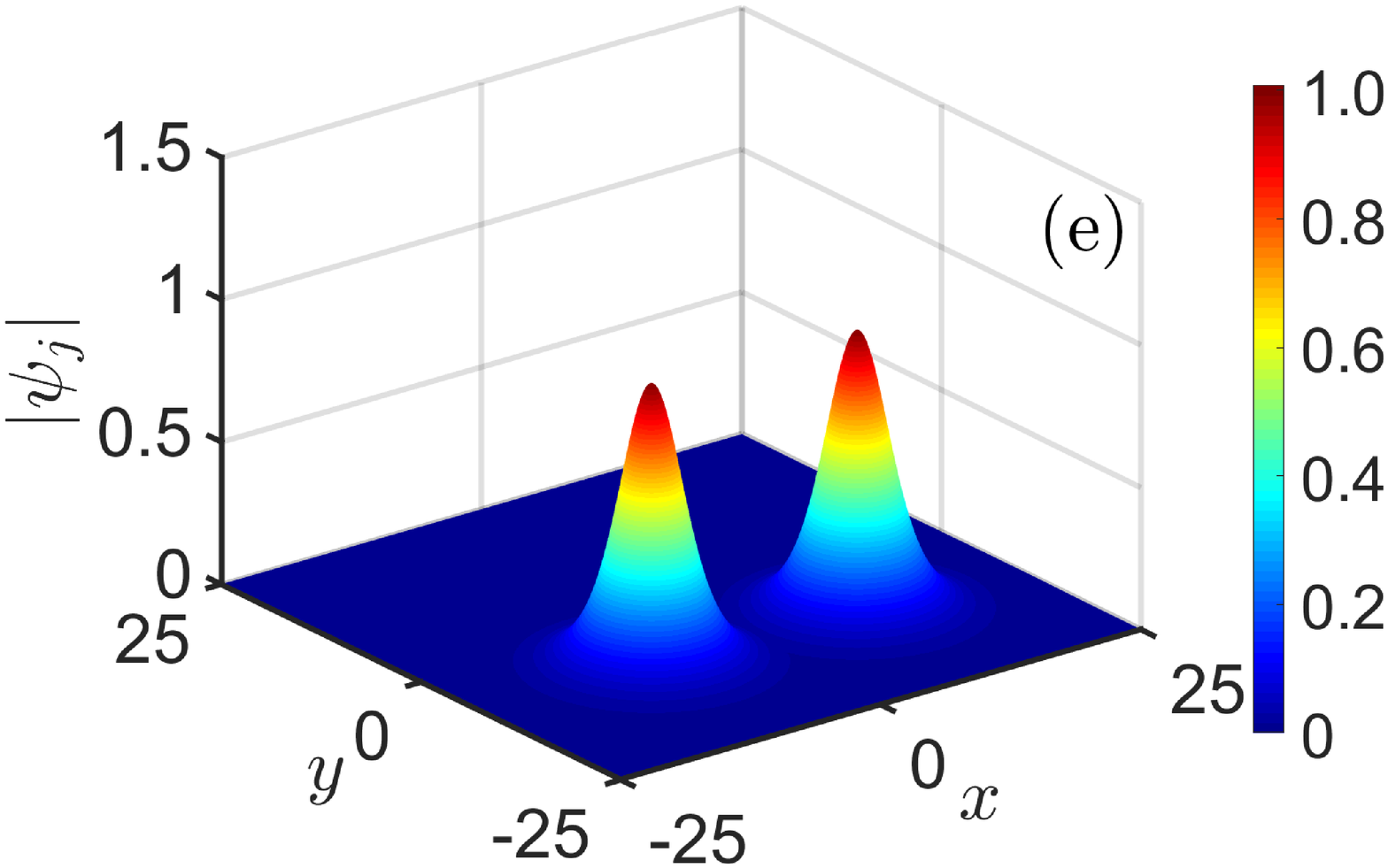} &
\epsfxsize=7cm  \epsffile{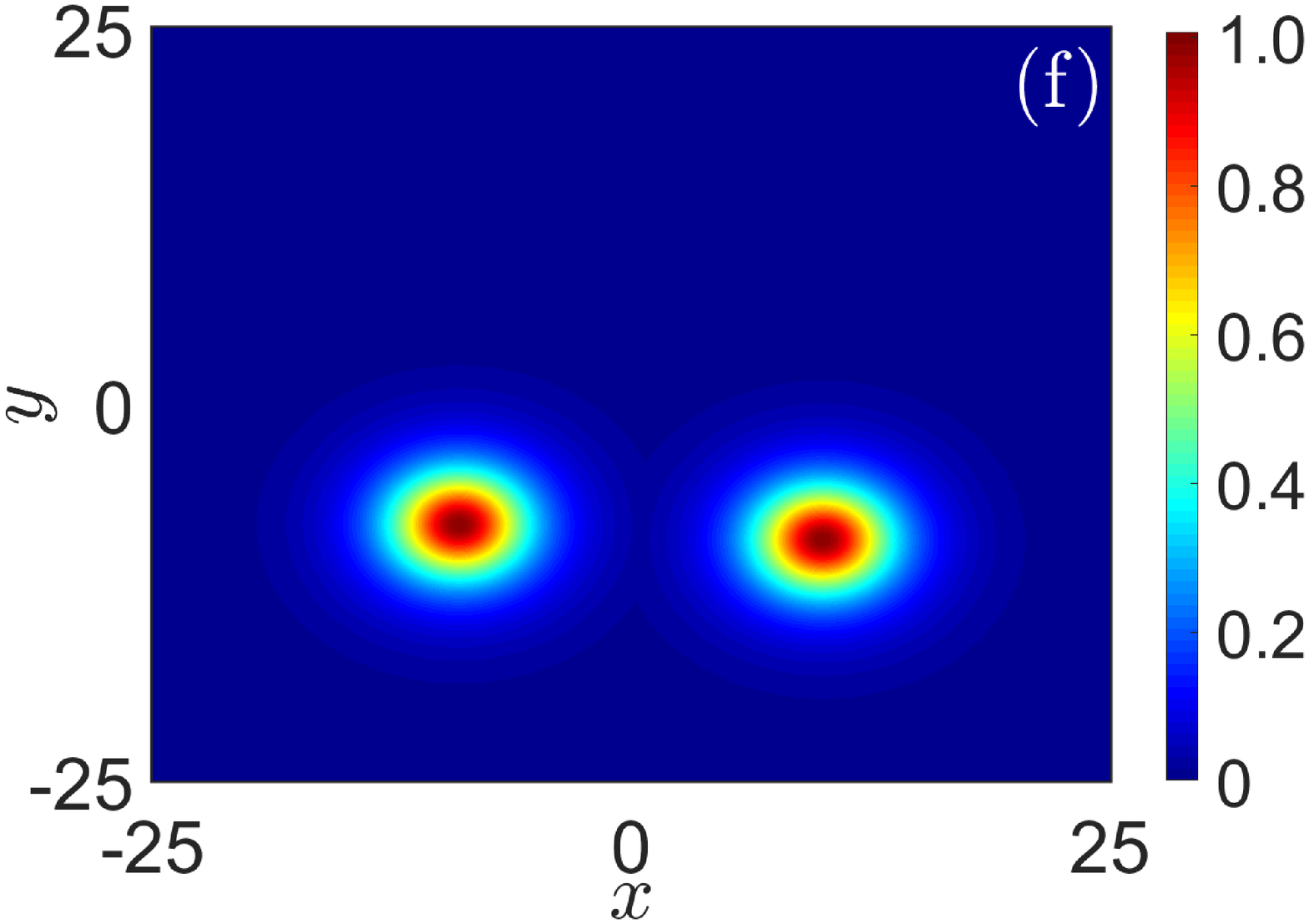} \\
\end{tabular}
\caption{(Color online) The initial soliton profiles (a) and the soliton profiles at $z=z_{i}=0.6$ (c) and at $z=z_{f}=1$ (e) in a two-soliton collision obtained by the simulation of equation (\ref{NLS_coll1}) with $\epsilon_{3}=0.01$ and $d_{11}=20$.
(b, d, f) The soliton profiles $|\psi_{j}(x, y, z)|$ of (a, c, e) by using the level colormap, respectively.
}
\label{fig1}
\end{figure}

Second, let us illustrate the collision between two solitons by the simulation of equation (\ref{NLS_coll1}) with cubic loss. We emphasize that the illustration with quintic loss is similar.
The parameters are: $\epsilon_{3}=0.01$, $(x_{10},y_{10})=(-10, 9)$, $(x_{20},y_{20})=(9, 8)$, $\textbf{d}_{1} = (20, -18)$, and $\textbf{d}_{2} = (-18, 16)$. Two velocity vectors $\textbf{d}_{1}$ and $\textbf{d}_{2}$ satisfy the relation (\ref{NLS_num1}) with $z_{c}=0.5$.
One can measure $\cos\theta = -0.1111$ and
$|\textbf{d}| = 38.0526$.
Figure \ref{fig1}(a) represents the initial soliton profiles $|\psi_{j}(x,y,0)|$ 
while figures \ref{fig1}(c) and (e) depict the soliton profiles $|\psi_{j}(x,y,z)|$, which are obtained by the simulation, at the intermediate distance $z_{i}=0.6>z_{c}=0.5$, as an example, and at the final distance $z_{f}=1$, respectively.
Figures \ref{fig1}(b, d, f) represent the soliton profiles $|\psi_{j}(x,y,z)|$ in form of the level colormap corresponding to the soliton profiles in figures \ref{fig1}(a, c, e), respectively. The agreement between the analytic predictions and the simulation results is very good. In fact, the relative errors in measuring $\Delta A_{1}^{(c)}$ and in measuring the soliton patterns at $z=z_{f}$ are 0.02 and 0.009, respectively.

\begin{figure}[ptb]
\begin{tabular}{cc}
\epsfxsize=7cm  \epsffile{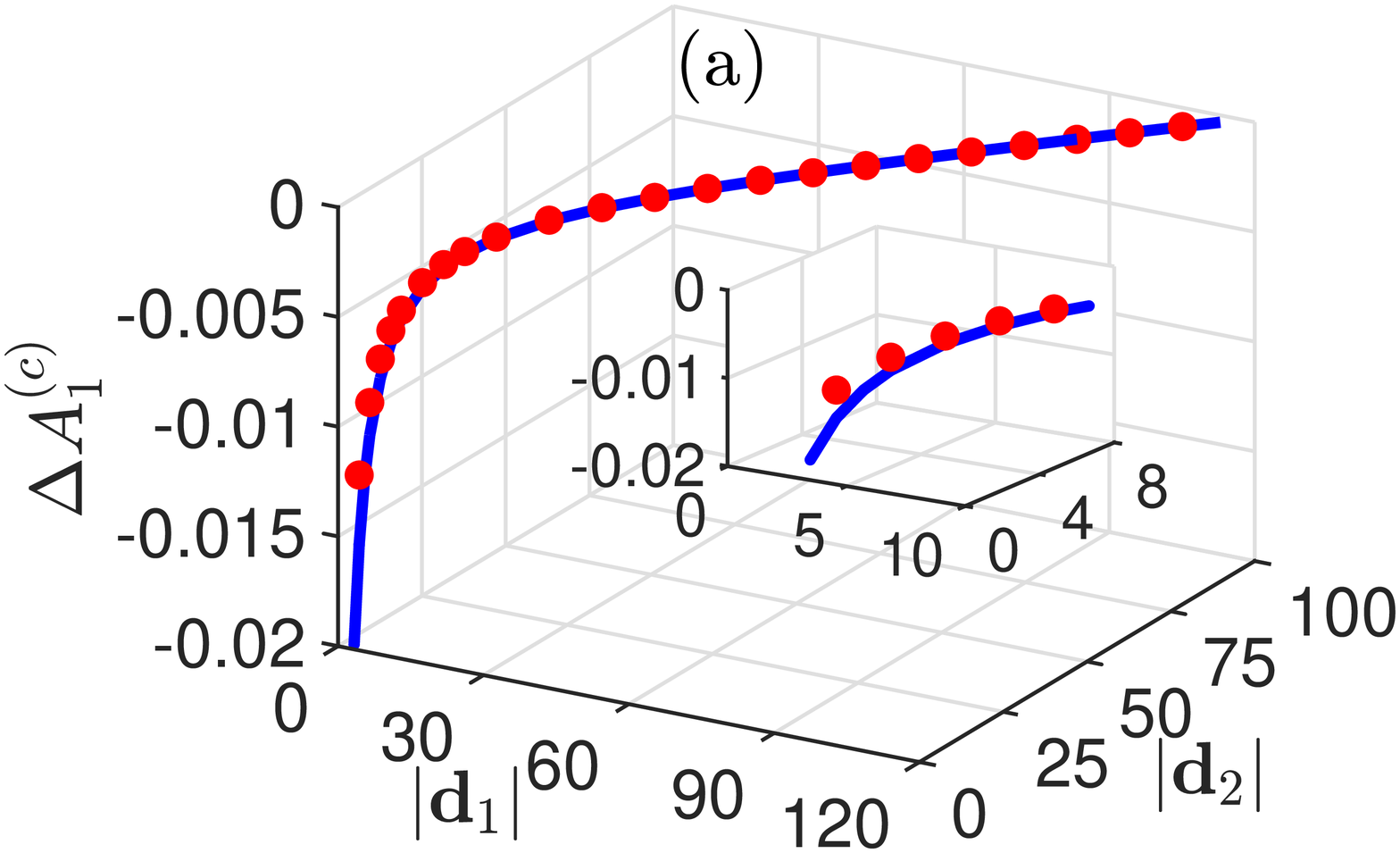} &
\epsfxsize=7cm  \epsffile{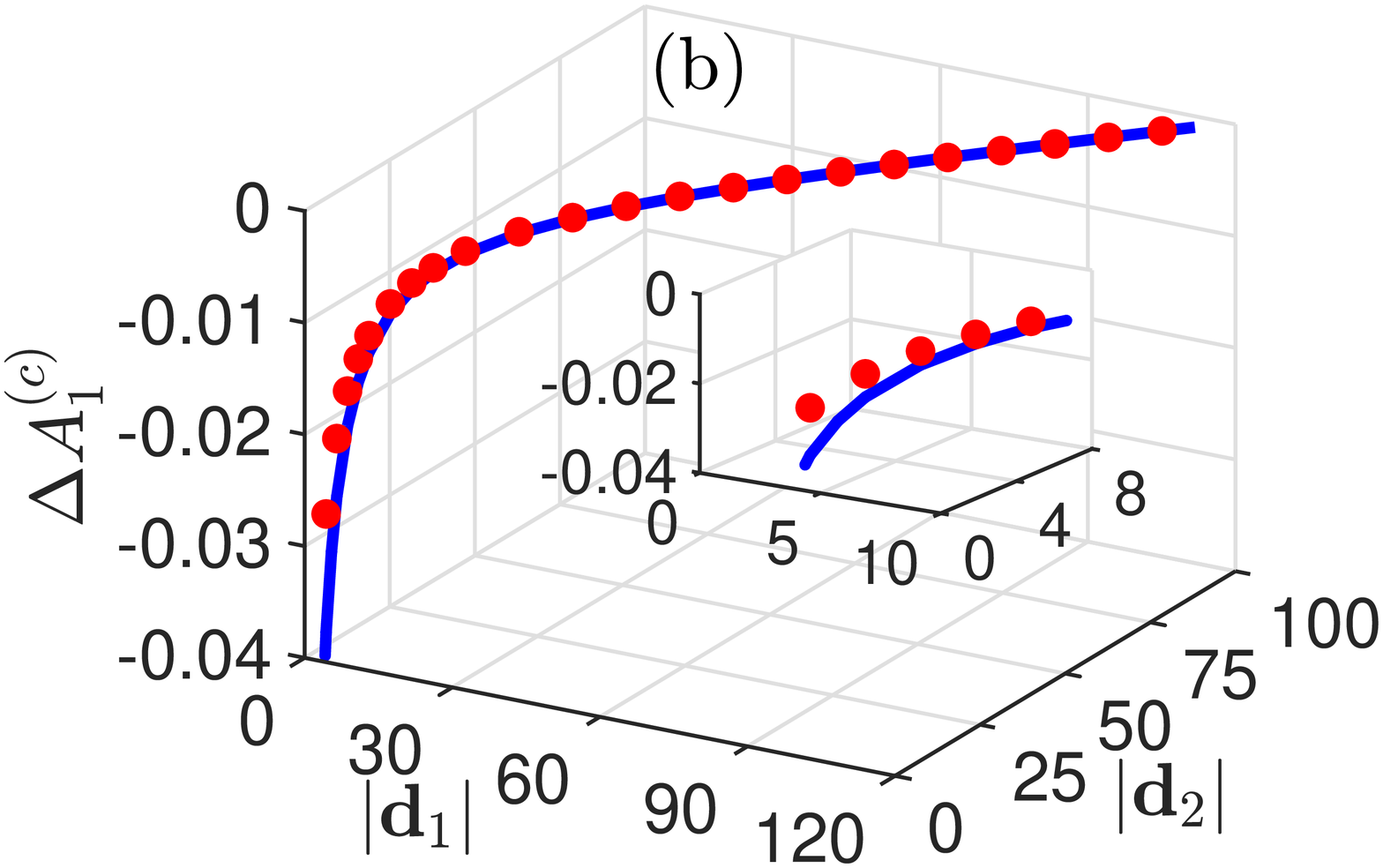} \\
\end{tabular}
\caption{(Color online) The dependence of $\Delta A_{1}^{(c)}$ on $\textbf{d}_{1}$ and $\textbf{d}_{2}$ with $\epsilon_{3}=0.01$ (a) and $\epsilon_{5}=0.01$ (b). The red circles correspond to the $\Delta A_{1}^{(c)}$ obtained by simulations of equation (\ref{NLS_coll1}). The solid blue curves represent the analytic prediction $\Delta A_{1}^{(c)}$ of equation (\ref{NLS_coll9}) for $m=1$ (a) and of equation (\ref{NLS_coll10}) for $m=2$ (b).
The inset of each figure represents $\Delta A_{1}^{(c)}$ at the small values of $|\textbf{d}_{1}|$ and $|\textbf{d}_{2}|$.
}
\label{fig2}
\end{figure}

\begin{figure}[ptb]
\begin{tabular}{cc}
\epsfxsize=7cm  \epsffile{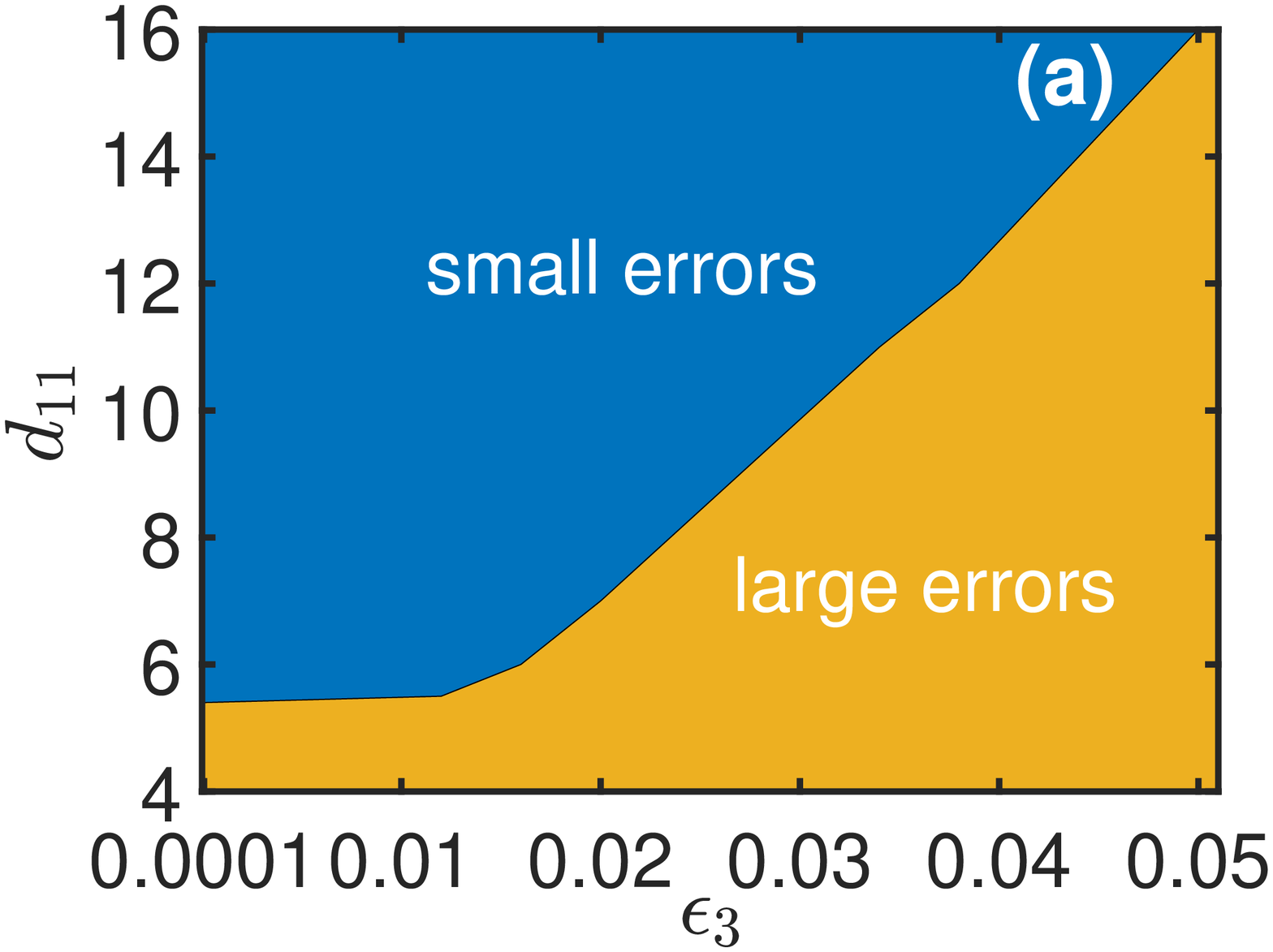} &
\epsfxsize=7cm  \epsffile{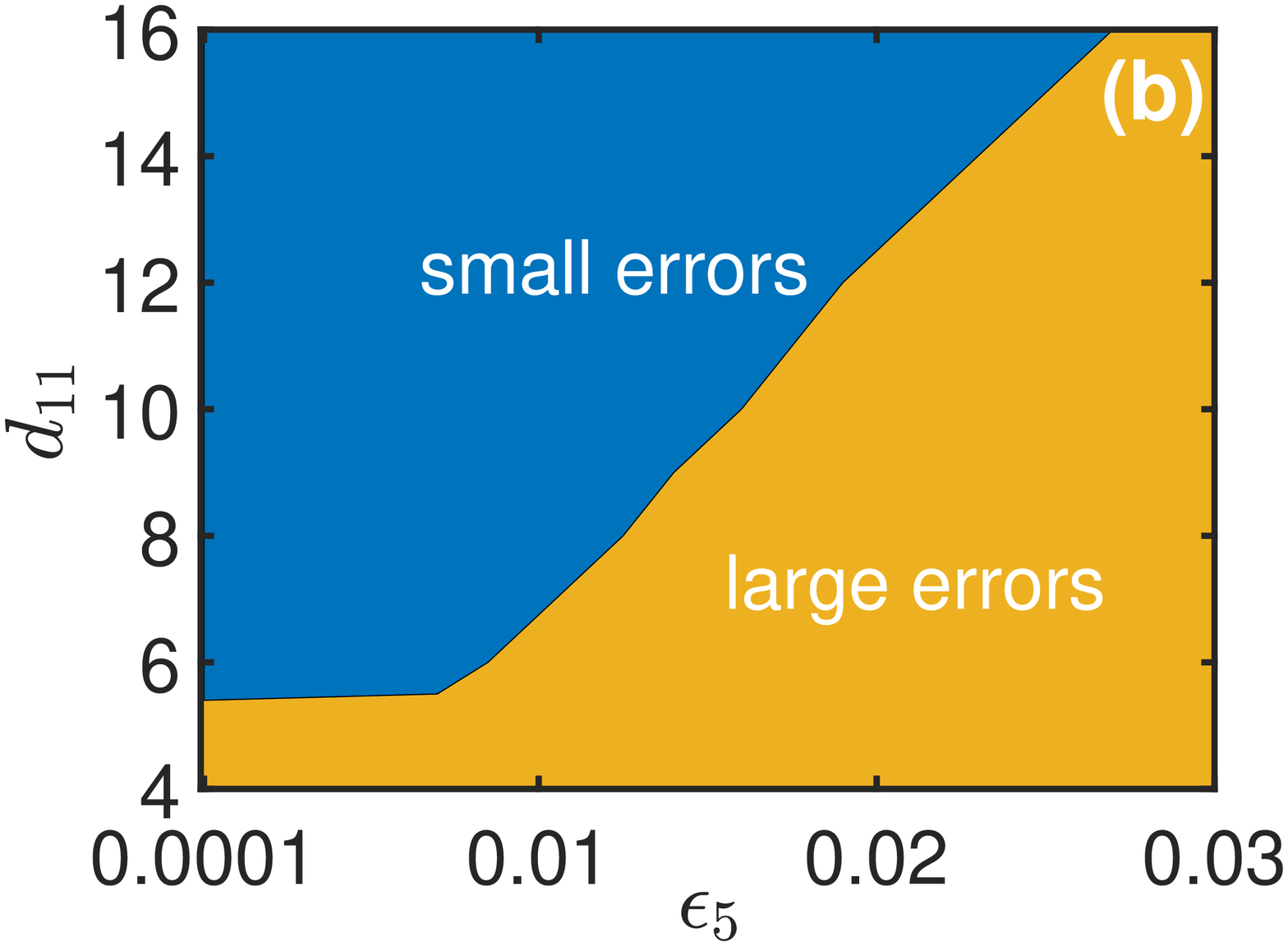} \\
\end{tabular}
\caption{(Color online) 
The simulation results of equation (\ref{NLS_coll1}) for a wide range values of $\epsilon_{2m+1}$ and $\textbf{d}_{j}$ with $m=1$ (a) and $m=2$ (b). The initial position parameters are $(x_{10},y_{10})=(-10, 9)$ and $(x_{20},y_{20})=(9, 8)$.
}
\label{fig5}
\end{figure}

Next, we study the dependence of $\Delta A_{1}^{(c)}$ on $\textbf{d}_{1}$ and $\textbf{d}_{2}$.
In simulations, the magnitudes of $\textbf{d}_{1}$ and $\textbf{d}_{2}$ will be changed while the value of $\theta$ is constant. The parameters are: $(x_{10},y_{10})=(-10, 9)$, $(x_{20},y_{20})=(9, 8)$, $\textbf{d}_{1} = (d_{11}, -0.9d_{11})$, $ \textbf{d}_{2} = (-0.9d_{11}, -0.8d_{11})$, where $2\le d_{11}\le 80$, and $z_{f}=2z_{c}$.
The velocities $\textbf{d}_{1}$ and $\textbf{d}_{2}$ satisfy the relation (\ref{NLS_num1}) with $z_{c}=10/d_{11}$.
One can measure $\cos\theta = -0.1111$ and $|\textbf{d}| = 1.9026d_{11}$. 
The loss coefficients are $\epsilon_{2m+1}=0.01$ and $\epsilon_{2m+1}=0.02$ for $m=1$ and $m=2$. The relative errors in the approximation of $\Delta A_{1}^{(c)}$ are less than $0.31$ for $2 \le d_{11} < 10$ and less than $0.07$ for $10 \le d_{11} \le 80$ for $m=1$.
They are less than $0.39$ for $2 \le d_{11} < 12$ and less than $0.1$ for $12 \le d_{11} \le 80$ for $m=2$.
The dependence of $\Delta A_{1}^{(c)}$ on $\textbf{d}_{1}$ and $\textbf{d}_{2}$ is depicted
in figure \ref{fig2} with $\epsilon_{3}=0.01$ (a) and $\epsilon_{5}=0.01$ (b).
Figure \ref{fig5} shows the simulation results for a wide range values of $\epsilon_{2m+1}$ and $\textbf{d}_{j}$ with $m=1$ (a) and $m=2$ (b).
The blue domain corresponds to the small errors, i.e., the relative errors in measuring $\Delta A_{1}^{(c)}$ and in measuring the soliton patterns at $z=z_{f}$ are less than or equal to $E_{a}$ and $E_{s}$, respectively.
Besides that the orange domain depicts the large errors, i.e., the relative error in measuring $\Delta A_{1}^{(c)}$ is greater than $E_{a}$ or the relative error in measuring the soliton patterns at $z=z_{f}$ is greater than $E_{s}$.
Moreover,  by implementing the simulations with $0.0001 \le \epsilon_{2m+1} \le 0.05$ and 
$2 \le d_{11} \le 80$, one can observe the small errors for $12 \le d_{11} \le 80$ with $0.0001 \le \epsilon_{2m+1} \le 0.02$, i.e., for fast collisions with weak nonlinear loss.

\begin{figure}[ptb]
\begin{tabular}{cc}
\epsfxsize=7cm  \epsffile{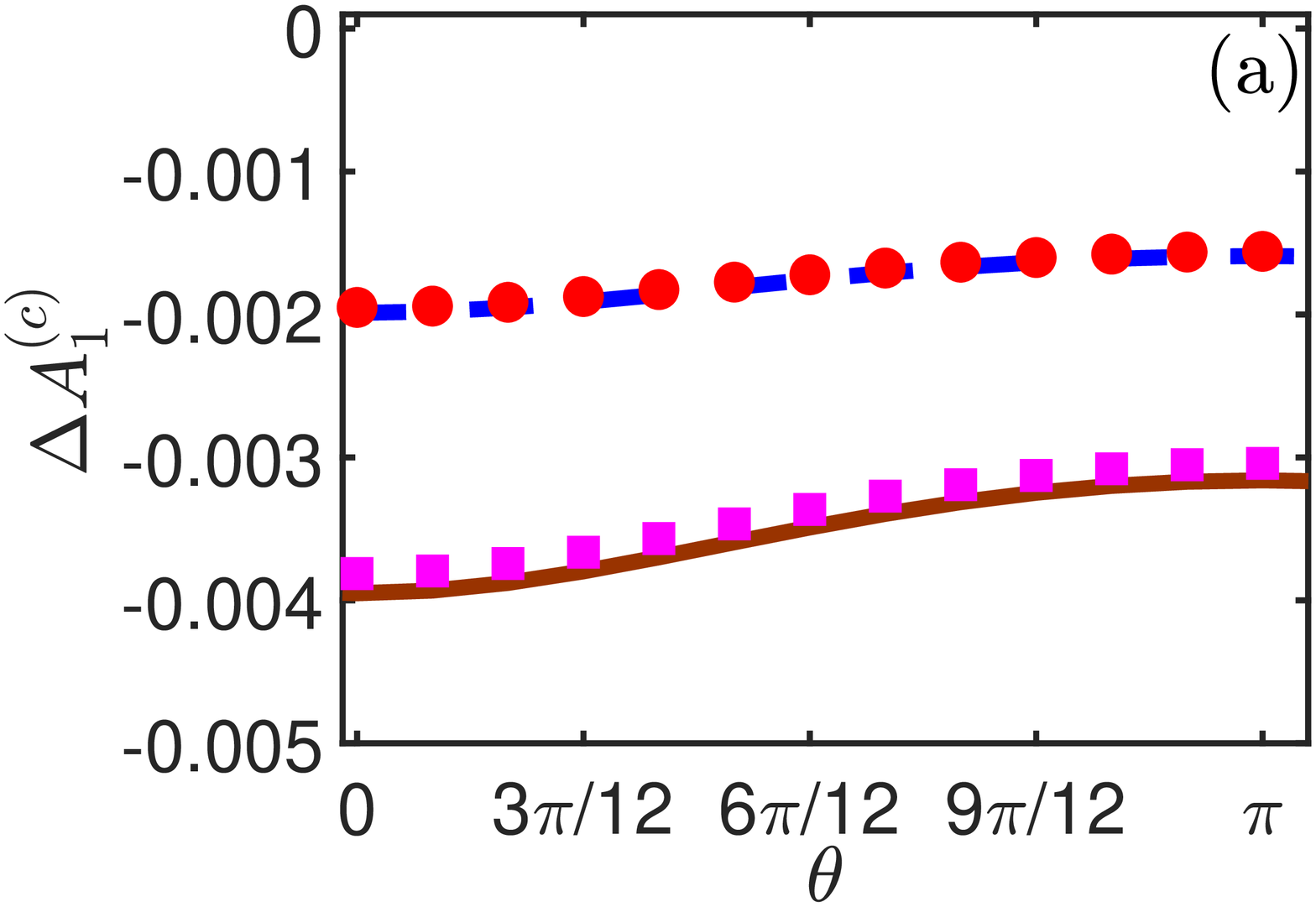} &
\epsfxsize=7cm  \epsffile{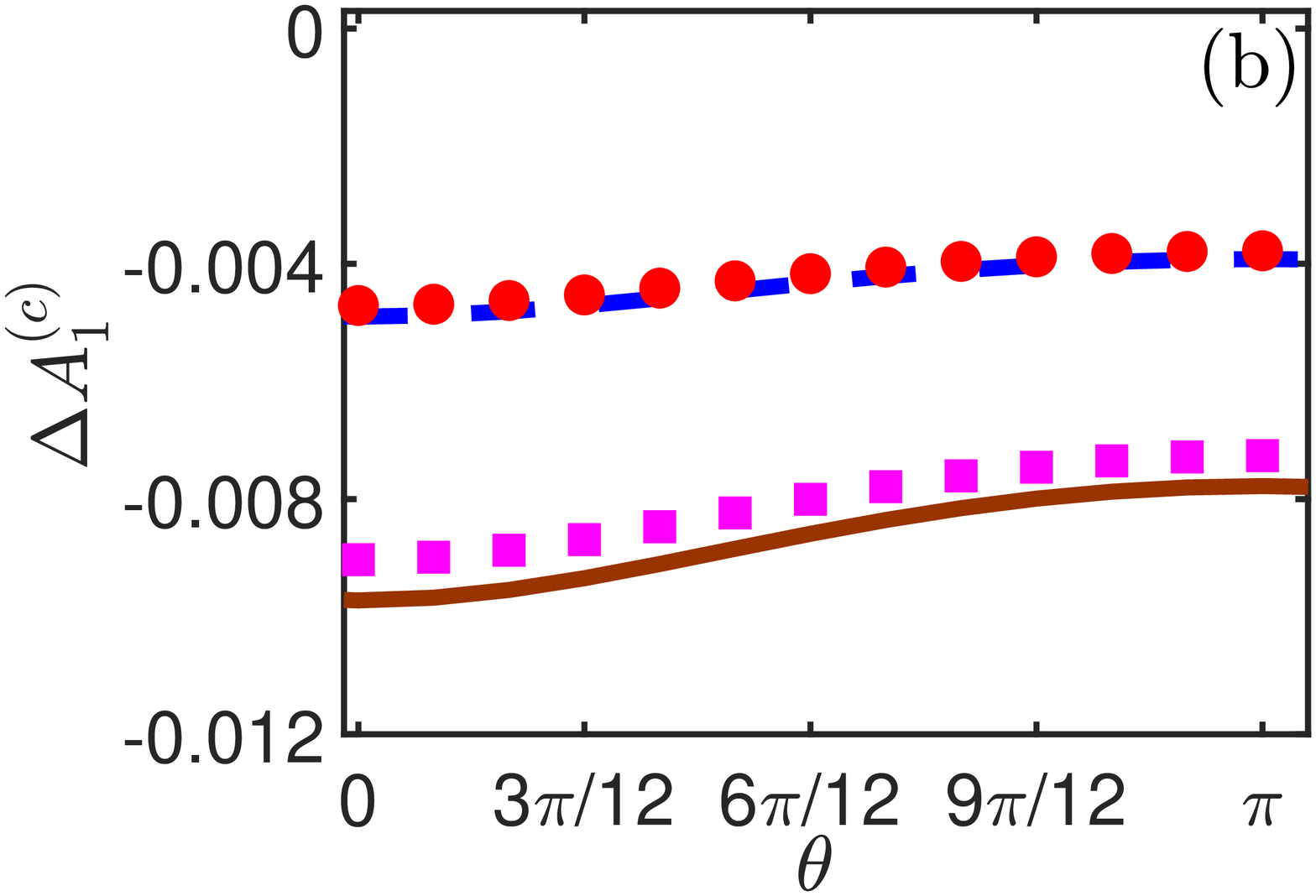} \\
\end{tabular}
\caption{(Color online) The dependence of $\Delta A_{1}^{(c)}$ on $\theta$ for $m=1$ (a) and $m=2$ (b). The red circles and purple squares correspond to $\Delta A_{1}^{(c)}$ obtained by simulations of equation (\ref{NLS_coll1}) with $\epsilon_{2m+1}=0.01$ and $\epsilon_{2m+1}=0.02$, respectively. The dashed blue and solid brown curves represent the analytic prediction $\Delta A_{1}^{(c)}$ with $\epsilon_{2m+1}=0.01$ and $\epsilon_{2m+1}=0.02$, respectively, of equation (\ref{NLS_coll9}) for $m=1$ and of equation (\ref{NLS_coll10}) for $m=2$.
}
\label{fig3}
\end{figure}

\begin{figure}[ptb]
\begin{tabular}{cc}
\epsfxsize=7cm  \epsffile{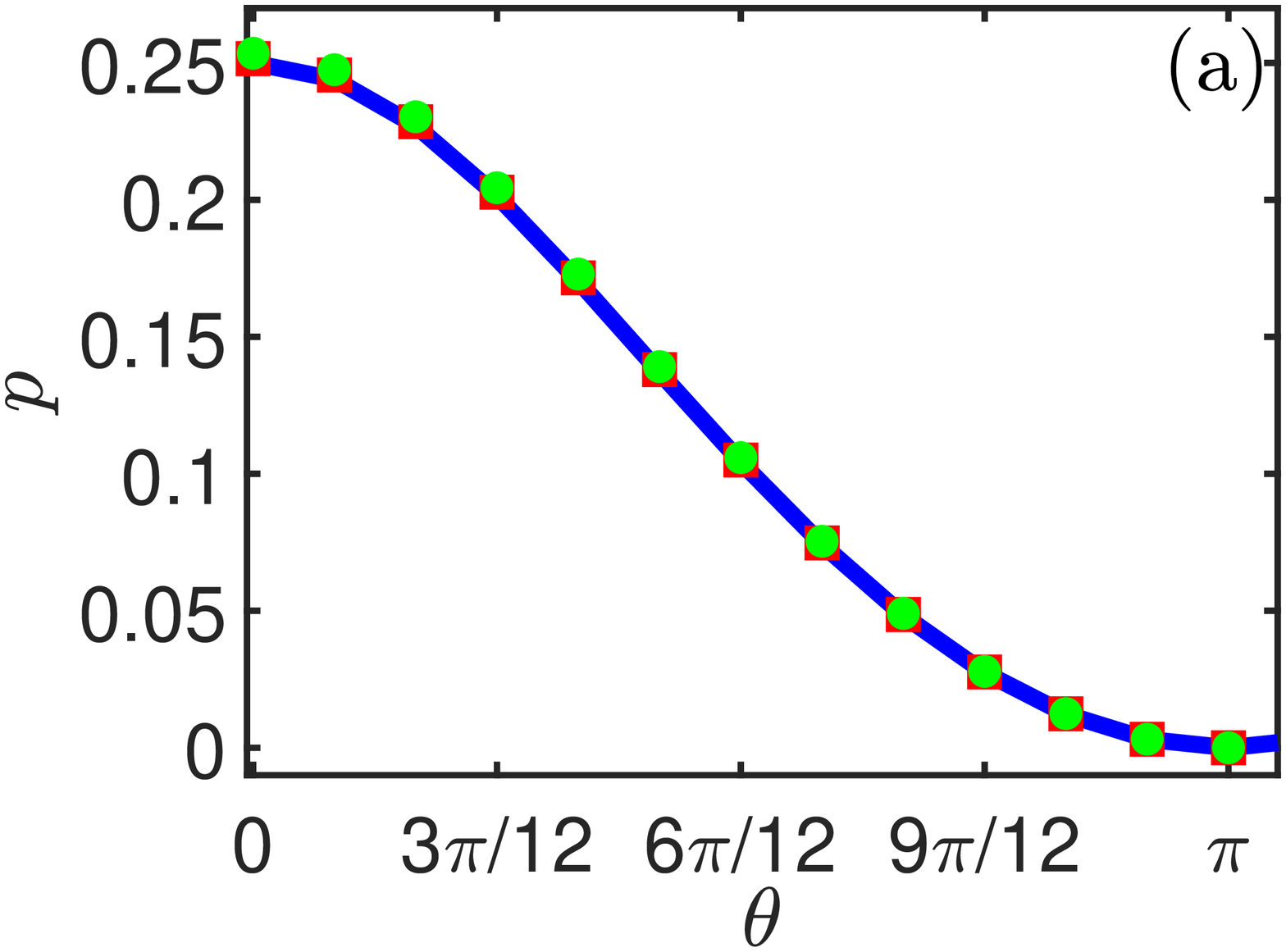} &
\epsfxsize=7cm  \epsffile{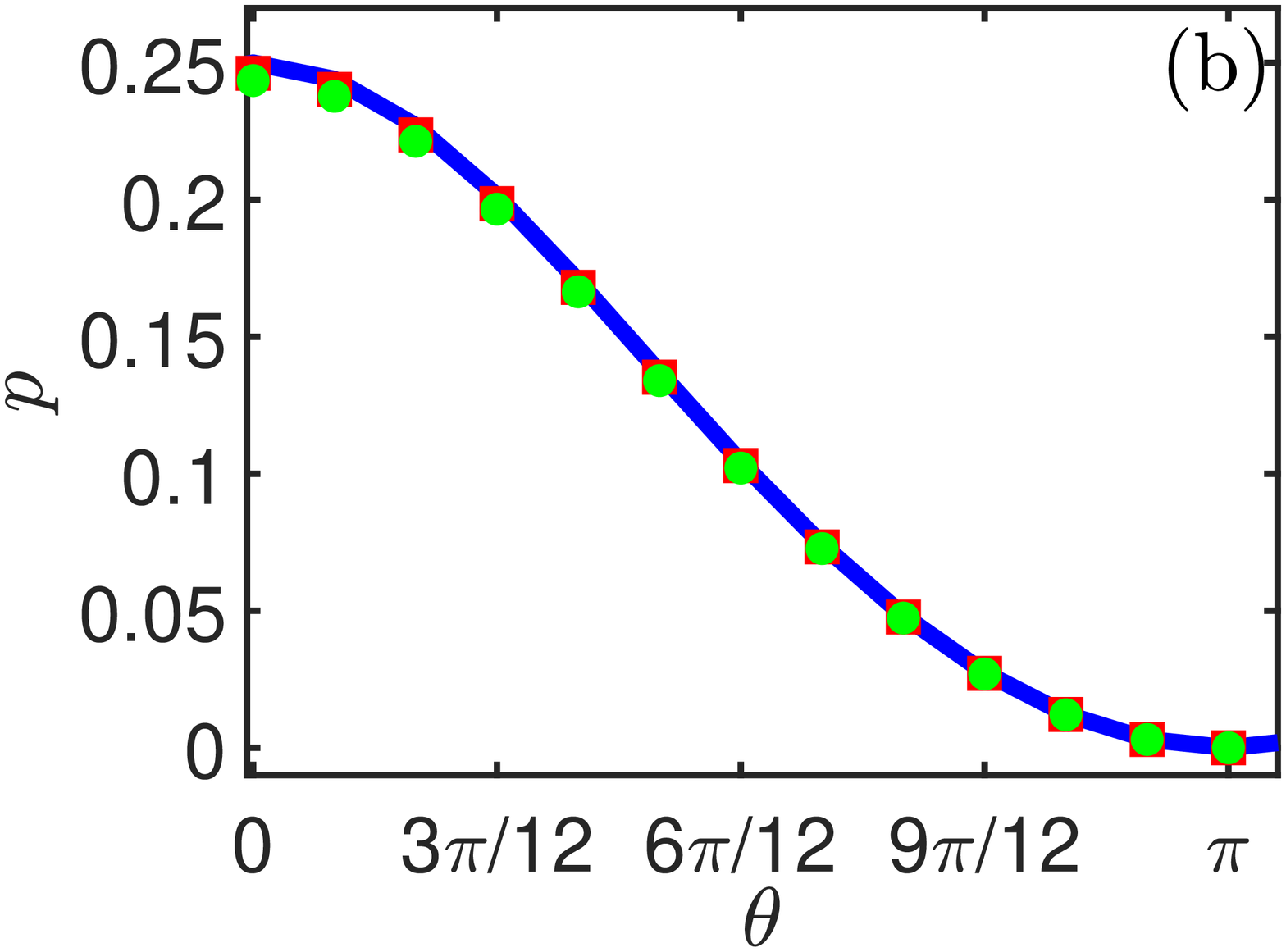} \\
\end{tabular}
\caption{(Color online) The dependence of $p$, which is measured from equation (\ref{NLS_relp}), on $\theta$ for $m=1$ (a) and $m=2$ (b). 
 The red squares and green circles represent the values of $p$ obtained by simulations of equation (\ref{NLS_coll1}) with $\epsilon_{2m+1}=0.01$ and $0.02$, respectively. The solid blue curves correspond to the theoretical prediction values of $p$.
}
\label{fig4_add}
\end{figure}

Finally, we describe the dependence of $\Delta A_{1}^{(c)}$ on $\theta$.
In simulations, the values of $\theta$ will be changed over the interval $[0,\pi]$ while $|\textbf{d}_{1}|$ and $|\textbf{d}_{2}|$ are constants.
For example, the parameters can be chosen:
$(x_{10}, y_{10})=\left(18\cos(9\pi/12), 18\sin(9\pi/12)\right)$,
$(x_{20}, y_{20})=\left(2\cos(k\pi/12), 2\sin(k\pi/12)\right)$,
$d_{11}=25$, $d_{12}=y_{10}d_{11}/x_{10}$, $d_{21}=x_{20}d_{11}/x_{10}$,
$d_{22}=y_{20}d_{11}/x_{10}$,
where $k=-3,-2,-1,0,1,2...,9$, and $z_{f}=2z_{c}$.
Two velocity vectors $\textbf{d}_{1}$ and $\textbf{d}_{2}$ satisfy the relation (\ref{NLS_num1}) with $z_{c}=0.5091$.
One can measure $|\textbf{d}_{1}| =35.3553$ and $| \textbf{d}_{2} | =3.9284$.
The agreement between the simulation results and the analytic predictions is very good.
In fact, the maximal relative errors in the approximation of $\Delta A_{1}^{(c)}$ over $[0, \pi]$ are $0.02$ and 0.038 for $\epsilon_{3}=0.01$ and $\epsilon_{3}=0.02$, respectively. They are $0.039$ for $\epsilon_{5}=0.01$ and 0.071 for $\epsilon_{5}=0.02$. 
Figure \ref{fig3} shows the dependence of $\Delta A_{1}^{(c)}$ on $\theta$ with $\epsilon_{2m+1}=0.01$ and $\epsilon_{2m+1}=0.02$ for $m=1$ (a) and $m=2$ (b). 
One can observe that the magnitude of $\Delta A_{1}^{(c)}$ is smaller for a larger value of $\theta$, i.e., for a faster collision.
Figure \ref{fig4_add} shows the dependence of the relative change $p$ on $\theta$ with $\epsilon_{2m+1}=0.01$ and $0.02$ for $m=1$ (a) and for $m=2$ (b). As can be seen, the relative difference $p$ is independent of the choices of $\epsilon_{2m+1}$ and $m$. The values of $p$ are decreasing from $p_{\max}=0.25$ at $\theta=0$ to $p_{\min}=0$ at $\theta=\pi$. The maximal relative error in calculations of $p$ over $[0,\pi]$ is 0.016 for $m=1$ and it is 0.026 for $m=2$.

In summary, the very good agreement between the analytic calculations for $\Delta A_{1}^{(c)}$ and the  simulation results of the perturbed coupled nonlinear Schr\"odinger model validated our theoretical calculations for $\Delta A_{1}^{(c)}$.

\section{Conclusions}
\label{conclusion}

We derived the expressions for the amplitude dynamics of 2D NLS solitons in a fast collision in the saturable nonlinear media with the generic weak ($2m + 1)-$order loss, for any $m \ge 1$. We first established the single soliton dynamics in the presence of the nonlinear loss.
Then, we derived the expressions for the collision-induced amplitude dynamics in a fast collision of two 2D solitons in the presence of the nonlinear loss. Our perturbative method is quite different to the traditional perturbative method derived by Kaup \cite{Kaup1990,Kaup1991}. The previous method was based on the projections of the total collision-induced change in the soliton envelope on the four localized eigenmodes of the linear operator $\hat{L}$ describing small perturbations about the fundamental 1D NLS soliton, where the unperturbed model is integrable and the {\it ideal} soliton solution was used \cite{Kaup1990,Kaup1991}.
In fact, in the current paper, the unperturbed (2+1)D NLS equation is nonintegrable. Our perturbative approach was based on the calculations on the energy balance of perturbed solitons, the analysis of the total collision-induced change in the soliton envelope, and the use of the {\it perturbed} single-soliton solution with the assumptions of weak nonlinear loss and fast collisions.
Consequently, the current method allows us to study the fast soliton collision-induced amplitude dynamics of fast 2D solitons of the nonintegrable model such as the (2+1)D NLS equations with a saturable nonlinearity.
The theoretical calculations were confirmed by extensive simulations of the corresponding coupled nonlinear Schr\"odinger models in the presence of the cubic loss $(m=1)$ and in the presence of the quintic loss $(m=2)$ with varying the loss coefficients and the velocity vectors. 

Our current perturbative approach can be applied for studying the soliton collision-induced amplitude dynamics for a larger class of soliton equations, even in a nonintegrable system, with other dissipative perturbations in a similar manner. 
Furthermore, we showed that the current perturbative approach can be applied to simply calculate the collision-induced amplitude shift in a fast collision of two 1D solitons for a wider class of perturbed (1+1)D NLS equations in a straightforward manner. More specifically, 
we applied the current perturbative technique to
derive the expression for the collision-induced amplitude shift in a fast collision of
two 1D cubic NLS solitons in the presence of the delayed Raman response. 
This expression has been also derived in Ref. \cite{CP2005} based on the perturbation technique developed by Kaup \cite{Kaup1990,Kaup1991} for 1D NSL solitons.
We expect that these results can open a way to study the collision-induced dynamics of solitons in 2D or in higher dimensions in other types of materials and can be also applied to study the collision-induced dynamics of beams in Bose-Einstein condensates.

\section*{Funding}
This work is funded by the Vietnam National Foundation for Science and Technology Development (NAFOSTED) under Grant No. 107.99-2019.340.



\appendix

\section{The efficiency of the current perturbation method} 
\label{Appen_NLS2}

In this Appendix, we illustrate that the current perturbation method is robust and simple to study the collision-induced amplitude dynamics in {\it fast} collisions of solitons of perturbed NLS equations. More specifically, one can apply the current perturbative approach to simply derive the expression for the collision-induced amplitude shift in a fast collision of two 1D NLS solitons with delayed Raman response in a straightforward manner. This expression has been derived in Ref. \cite{Kumar1998} by the Taylor expansion and in Ref. \cite{CP2005} by the traditional perturbation technique developed by Kaup \cite{Kaup1990,Kaup1991} for 1D NLS solitons. 

For this purpose, we consider a fast collision between two solitons under the framework of coupled (1+1)D cubic NLS equations with the delayed Raman response \cite{CP2005,PNT2016}:
\begin{eqnarray}&&
\!\!\!\!\!\!\!\!\!\!\!\!\!\!
i\partial_{z}\psi_{j} + \partial^{2}_{t} \psi_{j} + 2|\psi_{j}|^{2}\psi_{j} + 4|\psi_{l}|^{2}\psi_{j}
\nonumber \\&&
=  -\epsilon_{R}\psi_{j}\partial_{t}|\psi_{j}|^{2} - \epsilon_{R}\psi_{j}\partial_{t}|\psi_{l}|^{2}
-\epsilon_{R}\psi_{l}\partial_{t}(\psi_{j}\psi_{l}^{*}),
 \!\!\!\!\!\!\!\!\!\!\!\!\!\!
\label{Appendix_C1}
\end{eqnarray}
where $\epsilon_{R}$ is the Raman coefficient, $0 < \epsilon_{R} \ll 1$, $1\le j,l\le 2$, and $j\ne l$. 
The first term on the RHS of equation (\ref{Appendix_C1}) describes the Raman-induced intra-pulse interaction while the second and third terms describe the Raman-induced inter-pulse interaction.
We note that the unperturbed NLS equation $i\partial_{z}\psi_{j} + \partial^{2}_{t} \psi_{j} + 2|\psi_{j}|^{2}\psi_{j} =0$ has the fundamental soliton solution 
$\psi_{cs,j}(t,z) = \Psi_{cs,j}\exp(i\chi_{j})$,
where
\begin{eqnarray}&&
\Psi_{cs,j}(t,z) = \frac{\eta_{j}}{\cosh(x_{j})},
 \!\!\!\!\!\!\!\!\!\!\!\!\!\!
\label{Appendix_C2}
\end{eqnarray}
$x_{j}=\eta_{j}(t-y_{j} - 2\beta_{j}z)$, $\chi_{j}=\alpha_{j} + \beta_{j}(t-y_{j}) + (\eta_{j}^{2} - \beta_{j}^{2})z$, and parameters $\eta_{j}$, $\beta_{j}$, $\alpha_{j}$, and $y_{j}$ are related to the amplitude, frequency, phase, and position of the soliton $j$, respectively.
Similarly to Ref. \cite{CP2005}, we assume that $1/|\beta| \ll 1$ with $\beta=\beta_{2} - \beta_{1}$ and that two solitons are well separated at the initial propagation distance $z=0$ and at the final distance $z=z_{f}$.
One can look for the solution of equation (\ref{Appendix_C1}) in the form $\psi_{c,j}(t,z)=\psi_{c,j0}(t,z) + \phi_{c,j}(t,z)$, 
where $\psi_{c,j0}(t,z) = \Psi_{c,j0}(x_{j0})\exp(i\chi_{j0})$ is the single-soliton propagation solution of equation (\ref{Appendix_C1}) in the absence of inter-pulse interaction terms, i.e.,  $\psi_{c,j0}$ satisfies the following equation:
\begin{eqnarray}&&
\!\!\!\!\!\!\!\!\!\!\!\!\!\!
i\partial_{z}\psi_{j} + \partial^{2}_{t} \psi_{j} + 2|\psi_{j}|^{2}\psi_{j} 
=  -\epsilon_{R}\psi_{j}\partial_{t}|\psi_{j}|^{2},
 \!\!\!\!\!\!\!\!\!\!\!\!\!\!
\label{Appendix_C2b}
\end{eqnarray}
and $\phi_{c,j}(t,z)=\Phi_{c,j}(x_{j0})\exp(i\chi_{j0})$ represents a small correction to $\psi_{c,j0}$ due to inter-pulse interactions.

We now apply the current perturbation technique to calculate the collision-induced amplitude shift in a fast collision of two solitons described by equation (\ref{Appendix_C1}).
We first perform the calculations for the energy balance of equation (\ref{Appendix_C1}). It implies:
\begin{eqnarray}&&
\!\!\!\!\!\!\!\!\!\!\!\!\!\!
i\partial_{z}\int_{-\infty}^{\infty}|\psi_{j}|^{2}dt =
-\epsilon_{R}\int_{-\infty}^{\infty}\psi_{l}\psi_{j}^{*}\partial_{t}(\psi_{j}\psi_{l}^{*})dt
+ \epsilon_{R}\int_{-\infty}^{\infty}\psi_{l}^{*}\psi_{j}\partial_{t}(\psi_{j}^{*}\psi_{l})dt.
 \!\!\!\!\!\!\!\!\!\!\!\!\!\!
\label{Appendix_C3}
\end{eqnarray}
Equation (\ref{Appendix_C3}) can be written as:
\begin{eqnarray}&&
\!\!\!\!\!\!\!\!\!\!\!\!\!\!
i\partial_{z}\int_{-\infty}^{\infty}|\psi_{j}|^{2}dt =
-\epsilon_{R}\int_{-\infty}^{\infty}|\psi_{j}|^{2}\left(C_{l} - C_{l}^{*}\right)dt
-\epsilon_{R}\int_{-\infty}^{\infty}|\psi_{l}|^{2}\left(C_{j}^{*} - C_{j}\right)dt,
 \!\!\!\!\!\!\!\!\!\!\!\!\!\!
\label{Appendix_C4}
\end{eqnarray}
where $C_{k}=\psi_{k}\partial_{t}(\psi_{k}^{*})$ with $k=l,j$. By the definition of $\psi_{c,k}$, it implies
$C_{k}=\Psi_{c,k}\partial_{t}\Psi_{c,k} -i\beta_{k}\Psi_{c,k}^{2}$, where $\Psi_{c,k} = \Psi_{c,k0} + \Phi_{c,k}$.
Substituting the relation for $C_{k}$ into equation (\ref{Appendix_C4})
and using the adiabatic perturbation theory for the NLS soliton with concentrating only on the leading-order effects of the collision, it implies the energy balance equation for soliton 1: 
\begin{eqnarray}&&
\!\!\!\!\!\!\!\!\!\!\!\!\!\!
\partial_{z}\int_{-\infty}^{\infty}\Psi_{c,1}^{2}dt =
2\epsilon_{R}\beta\int_{-\infty}^{\infty}\Psi_{c,10}^{2}\Psi_{c,20}^{2}dt.
 \!\!\!\!\!\!\!\!\!\!\!\!\!\!
\label{Appendix_C5}
\end{eqnarray}
Integrating equation (\ref{Appendix_C5}) with respect to $z$ over the collision-interval $[z_{c}-\Delta z_{c}, z_{c}+\Delta z_{c}]$, it yields
\begin{eqnarray}&&
\!\!\!\!\!\!\!\!\!\!\!\!\!\!
\int_{z_{c}-\Delta z_{c}}^{z_{c}+\Delta z_{c}}\partial_{z}\int_{-\infty}^{\infty}\Psi_{c,1}^{2}dtdz=2\epsilon_{R}\beta M,
 \!\!\!\!\!\!\!\!\!\!\!\!\!\!
\label{Appendix_C6}
\end{eqnarray}
where $M=\int_{z_{c}-\Delta z_{c}}^{z_{c}+\Delta z_{c}}\int_{-\infty}^{\infty}\Psi_{c,10}^{2}(t,z)\Psi_{c,20}^{2}(t,z)dtdz$. By the definition of $\psi_{c,1}(t,z)$ and $\psi_{c,10}(t,z)$, one can use the approximation $\psi_{c,1}(t,z_{c}^{-})\simeq \psi_{c,10}(t,z_{c}^{-})$. Equation (\ref{Appendix_C6}) then leads to
\begin{eqnarray}&&
\!\!\!\!\!\!\!\!\!\!\!\!\!\!
\int_{-\infty}^{\infty}\Psi_{c,1}^{2}(t,z_{c}^{+})dt - \int_{-\infty}^{\infty}\Psi_{c,10}^{2}(t,z_{c}^{-})dt = 2\epsilon_{R}\beta M.
 \!\!\!\!\!\!\!\!\!\!\!\!\!\!
\label{Appendix_C7}
\end{eqnarray}
We note that
\begin{eqnarray}&&
\!\!\!\!\!\!\!\!\!\!\!\!\!\!
\int_{-\infty}^{\infty}\Psi_{c,j0}^{2}(t,z)dt =2\eta_{j0}(z),
\!\!\!\!\!\!\!\!\!\!\!\!\!\!
\label{Appendix_C8}
\end{eqnarray}
where $\eta_{j0}(z)$ is the amplitude parameter of $\psi_{c,j0}$ in the presence of the delayed Raman respone. By using the standard adiabatic perturbation theory, one can obtain $\eta_{j0}(z)=\eta_{j0}(0)$.
On the other hand, 
$\Psi_{c,1}(t,z_{c}^{+})$ can be expressed in the following manner:
\begin{eqnarray}&&
\!\!\!\!\!\!\!\!\!\!\!\!\!\!
\int_{-\infty}^{\infty}\Psi_{c,1}^{2}(t,z_{c}^{+})dt = 
\int_{-\infty}^{\infty} \frac{\eta_{1}^{2}(z_{c}^{+})}{\cosh^{2} \left[x_{1}(t,z_{c}^{+}) \right]}dt
=2\eta_{1}(z_{c}^{+}),
\!\!\!\!\!\!\!\!\!\!\!\!\!\!
\label{Appendix_C9}
\end{eqnarray}
where $ \eta_{1}(z_{c}^{+}) \simeq \eta_{1}(z_{c}^{-}) + \Delta \eta_{1}^{(c)} \simeq \eta_{10}(z_{c}^{-}) + \Delta \eta_{1}^{(c)}$ and $\Delta \eta_{1}^{(c)}$ is the total collision-induced amplitude shift of soliton 1.
Next, we calculate the integral $M$ by using the algebra approximations which were used to calculate the integral $M_{k,m}$ in equation (\ref{NLS_coll7}).
That is, one can take into account only the fast dependence of $\Psi_{c,j0}$ on $z$, which is the factor $v_{j} = t- y_{j} - 2\beta_{j}z$, and approximate other slow varying terms over $[z_{c} - \Delta z_{c}, z_{c} + \Delta z_{c}]$ by their values at $z_c$. This approximation of $\Psi_{c,j0}(t,z)$ is denoted by $\bar \Psi_{c,j0}(v_{j},z_{c})$. Moreover, since the integrand of $M$ is sharply peaked at a small interval $[z_{c} - \Delta z_{c}, z_{c} + \Delta z_{c}]$ about $z_{c}$, the limits of the integral $M$ can be extended to $-\infty$ and $\infty$. 
By changing the integration variable with $v_{j}=t-y_{j}-2\beta_{j}z$, it implies
$$
M=\frac{1}{2|\beta|}\int_{-\infty}^{\infty}\int_{-\infty}^{\infty}
\bar\Psi_{c,20}^{2}(v_{2},z_{c})\bar\Psi_{c,10}^{2}(v_{1},z_{c})dv_{1}dv_{2}.
\!\!\!\!\!\!\!\!\!\!\!\!\!\!
$$
Note that 
$$\int_{-\infty}^{\infty}\bar\Psi_{c,j0}^{2}(v_{j},z_{c})dv_{j} = \int_{-\infty}^{\infty}\Psi_{c,j0}^{2}(t,z_{c})dt.$$
It leads to:
\begin{eqnarray}&&
\!\!\!\!\!\!\!\!\!\!\!\!\!\!
M=\frac{1}{2|\beta|}
\int_{-\infty}^{\infty}\Psi_{c,20}^{2}(t,z_{c})dt\int_{-\infty}^{\infty}\Psi_{c,10}^{2}(t,z_{c})dt.
\!\!\!\!\!\!\!\!\!\!\!\!\!\!
\label{Appendix_C10}
\end{eqnarray}
Substituting equations (\ref{Appendix_C8}), (\ref{Appendix_C9}), and (\ref{Appendix_C10}) into equation (\ref{Appendix_C8}), it arrives at the equation for energy exchange:
\begin{eqnarray}&&
\!\!\!\!\!\!\!\!\!\!\!\!\!\!
\Delta \eta_{1}^{(c)} = 2\epsilon_{R}\sgn(\beta)\eta_{10}\eta_{20}.
\!\!\!\!\!\!\!\!\!\!\!\!\!\!
\label{Appendix_C11}
\end{eqnarray}
Equation (\ref{Appendix_C11}) is in the same form with equation (20) in Ref. \cite{CP2005} which was originally based on the perturbation technique developed by Kaup for 1D NLS solitons \cite{Kaup1990,Kaup1991}. In addition, we note that one can apply the current perturbative approach to simply derive the expression for the collision-induced amplitude shift in a fast collision of two 1D NLS solitons with nonlinear loss in a straightforward manner.

{}


\begin{thebibliography}{}
\bibitem{dimensionless} The dimensionless distance $z$ is $z=Z/L_{d}$, where $Z$ is the dimensional distance, $L_{d} = \beta_{0}w_{0}^{2}$ is the diffraction length, $w_{0}$ is the transverse scaling parameter related to the input beam width, and $\beta_{0}$ is the propagation constant in terms of the optical wavelength. The dimensionless coordinates $x$ and $y$ are $x=X/w_{0}$ and $y=Y/w_{0}$, where $X$ and $Y$ are the dimensional coordinates.
The coefficients $\epsilon_{2m+1}$ are related to the dimensional nonlinear loss $\rho_{2m+1}$ by $\epsilon_{2m+1} = 2\rho_{2m+1}P_{0}/\tilde \gamma$, where $P_{0}$ is the peak power and $\tilde \gamma$ is related to the nonlinearity coefficient of the nonlinear material.
\end{thebibliography}
\end{document}